# Protocellular energetics: Free energy estimates for all metabolic, self-assembly and vesicle fission processes

*Steen Rasmussen*[1,2,*], *Thomas Frederiksen*[3,4], *Masayuki Imai*[5], *Sheref S. Mansy*[6], *Sabine Muller*[7], *Marek Grzelcza*[8,3]

[1]Department for Physics, Chemistry and Pharmacy, University of Southern Denmark (SDU)
[2]Santa Fe Institute (SFI), New Mexico, USA; *steensantafe@gmail.com
[3]Donostia International Physics Center (DIPC), San Sebatian, Spain
[4]Ikerbasque, Basque Foundation for Science, Bilbao, Spain; thomas_frederiksen@ehu.eus
[5]Department of Physics, Tohoku University, Japan; imai@bio.phys.tohoku.ac.jp
[6]Department of Cellular, Computational and Integrative Biology (CIBIO) Trento, Italy; sheref.mansy@unitn.it
[7]Institute for Biochemistry, Greifswald University, Germany; sabine.mueller@uni-greifswald.de
[8]Materials Physics Center (MFC), San Sebastian, Spain; grzelczak.marek@gmail.com

*Point of contact

## Abstract

As minimal cells or protocells are dramatically simpler than modern unicells it is possible to quantitatively estimate free energy changes for every process in the lifecycle of a protocell and compare these with estimates of the free energy changes for lifecycles in modern unicells. We present quantitative estimates of all metabolic changes in part by new density function theory (DFT) estimations, in part by compiling previously measured or estimated free energy changes, and in part by new thermodynamic calculations for all self-assembly, vesicle bending, and fission energies.

# 1 Introduction

It is unclear how the first metabolisms emerged and operated in concert with information and containment at the origins of life, so we have developed a physicochemical model of a minimal living system that could provide insights about this question. The motivation for this work is to obtain a deeper understanding of the free energy changes for all reactions in a simple protocellular lifecycle as these free energy transformations are the driver for all living processes. It is possible to make a complete analysis due to the simplicity of a protocell opposed to the much more complex modern unicells, and having done such an analysis we can compare and contrast our findings to what is known about the free energy changes in modern unicell lifecycles.

We present a complete analysis of the free energy hierarchy with metabolic reactions that involves the formation and breakage of chemical bonds, component self-assembly processes, as well as the weak energies responsible for membrane bending and vesicle fission processes. To do this we've developed a methodological framework that also could be used for other simple protocellular or modern biological systems. Most previous protocellular lifecycle work focuses on component analysis, coupling of chemical mechanics, or reaction kinetics, and it does not include an analysis of a protocellular replication changes in system that consist of an integrated metabolic, information and container system. For recent related work, e.g. see Morrow et al (2019) where the authors nicely delineate the different kinds of free energy changes in their system before they do their own systems analysis, as well as Otto (2022) that also has summarized the free energy changes through a lifecycle of a life-like chemical system. The necessary energy required for synthesizing different modern and much more complex unicells, including the artificially synthesized JCVI-syn3A, are estimated and nicely summarized in Ortega-Arzola (2024).

Not including the top-down methods employed by Venter and others [Gibson et al (2010)], several groups are attempting to build protocells from component parts, some using the existing biological parts, including a modern translation machinery, while others use much simpler organic and inorganic building blocks. For example, there are laboratories that have expanded the range of materials to construct the compartment [Mu et al. (2024), Ianeselli et al. (2022), Abbas et al. (2021), Lee et al. (2024), Parrilla-Gutierrez et al. (2017), Madanan et al. (2024), Kahana & Lancet (2021), Cejkova et al. (2017)] elucidated mechanisms of growth and division of the compartment [Toparlak et al.(2023), Walde et al. (1994), Pearce et al. (2024)] optimized enzymatic [Sczepanski & Joyce (2014), Attwater et al. (2018), Cojocaru & Unrau (2021)] and nonenzymatic [Jia et al. (2024), Wu & Orgel (1992)] copying of RNA, and demonstrated the importance of out-of-equilibrium chemistry to fuel the system [Liu et al. (2025), Zhang et al. (2024), Harmsel et al. (2023), Zambrano et al. (2024), Matreux et al. (2024), Ashkenasy et al. (2023)] as well as more theoretical and computational investigations [Fellermann et al. (2015), Smith et al (2021), Sole et al. (2024), Arcas et al. (2024), Thomsen et al. (2025)]. A review of protocellular work can be found [Rasmussen et al (2008)] and a nice recent compilation of progress relevant for bottom-up approaches to minimal life can be found in the special issue edited by Sole et al. (2025).

# 2 Modern unicell energetics

Living cells and organisms must perform work to survive and reproduce. The ability to utilize and convert free energy into biological work is a fundamental characteristic of all living organisms. In general, free energy is defined as the capacity to perform work or bring about change. Thus, modern organisms carry out a remarkable variety of energy conversions. For example, chemical energy is used to drive the synthesis of complex macromolecules from simple precursors, or to generate concentration and electrical gradients of free energy for later use. It can also be transformed into movement, heat or light, as seen in the bioluminescence of some insects and deep-sea fish.

Biological organisms are open systems that exchange energy with their surroundings. They either capture energy from sunlight via photosynthesis, or extract energy from energy-rich molecules through metabolism, releasing heat as a by-product. Like all processes in the physical world, these energy conversions obey the laws of thermodynamics. In every metabolic process, as with all energy transformations, some of the usable energy (free enthalpy) is lost while unusable energy in the form of heat and entropy increases. Consequently, energy flows in one direction through the biosphere, and organisms cannot reclaim usable energy from energy that has dissipated as heat. Therefore, the key challenge for all living organisms is to capture energy from their environment in a form that can be transformed into usable energy for biological processes. Living cells have evolved mechanisms to meet this challenge. Chemical energy stored in organic molecules, i.e. sugars and fats, is transferred and transformed through a series of biochemical reactions into the high-energy molecule adenosine triphosphate (ATP) [Skulachev (1992)]. Heterotrophic cells obtain energy from nutrient molecules, while photosynthetic cells capture energy from sunlight. Both cell types convert free enthalpy into ATP or other energy-rich compounds, which power cellular processes at a constant temperature.

ATP functions as one of the main energy currencies of the cell. This molecule temporarily stores and transports energy to drive endergonic reactions. When ATP is used, one of its phosphoanhydride bonds is broken, releasing energy and forming either ADP or AMP. The energy that is stored in these phosphanhydride bonds either comes from organic molecules (food/fuel) or sunlight. Few ATP molecules are made directly from the bond breaking steps of a fuel molecule during catabolic metabolism. Instead, many more ATP molecules are made by harvesting the electrons. For example, when glucose is broken down to pyruvate during glycolysis, only a net of two ATP molecules is made. In aerobic organisms, an additional ATP is generated directly from the next metabolic cycle, i.e. the citric acid or Krebs cycle. However, during the oxidative step of glycolysis, passage through the pyruvate dehydrogenase complex, and the four oxidative steps of the citric acid cycle, one glucose molecule generates 10 NADH (from NAD+) and 2 FADH2 (from FAD), each carrying two donatable electrons. The thermodynamically favorable transfer of these electrons from NADH and FADH2 to the electron transport chain (ETC) is used to deposit energy in the form of a proton gradient, which is exploited by ATP synthase to generate an additional ca. 28 ATP molecules. The terminal electron acceptor of the ETC is oxygen. If oxygen is unavailable, organisms may use nitrate, fumarate, sulfate, or elemental sulfur as a terminal electron acceptor. Alternatively, cells survive with much fewer ATP molecules by exploiting fermentation, a metabolic pathway that does not exploit the electrons carried by NADH or FADH2 for the generation of ATP.

Plants use water instead of organic molecules to donate electrons that are shuttled down an analogous ETC that similarly generates a pH gradient during the light-dependent reactions of photosynthesis. The energy stored in this pH gradient is used to synthesize ATP, as seen above. Light energy is captured by chlorophyll and other pigments, and this energy is necessary to drive the endergonic electron transfer steps, e.g. extracting an electron from water, of the ETC. The terminal electron acceptor here is NADP+, forming NADPH. NADP+ is a phosphorylated form of NAD+ that is used in anabolism rather than catabolism. In this case, NADP+ (along with ATP) is used in the subsequent steps of photosynthesis to convert carbon dioxide into glucose. Although both types of organisms rely on different energy sources, organic molecules vs. sunlight, both use similar strategies to efficiently transform these environmentally available energy sources into biologically usable energy.

To summarize: in a living cell available chemical energy is utilized to break and make chemical bonds both for fuel and for building blocks.

As an example, a single E. coli cell uses approximately 9.54 x $10^{-11}$ Joules of energy to synthesize all its cellular components [Ortega-Arzola et al. (2024)]. During exponential growth, the cell consumes around 6.38 million ATP molecules per second, translating to a power consumption of roughly 0.57 pW (5.7 x $10^{-13}$ J/s). This energy is used for various processes, including DNA replication, protein synthesis, and membrane synthesis. Assuming the sole resource input for the ATP production is glucose in an anaerobic environment, the free energy harvesting efficiency is above 50% [Darnell et al. (1986)]. This means that for each cellular reproduction cycle an E. coli per must harvest about 2 x $10^{-10}$ Joules from resource glucose.

As another example, the initial quantum yield of photosynthesis from the available photon energy in sunlight is about 30% (collected by chlorophyll), while the resulting energy efficiency for glucose production is about 5%, and the final biomass free energy efficiency is between 0.1 and 1%.

# 3. An example minimal protocell

To make a quantitative estimate of all free energy changes in a lifecycle, we need to pick a concrete protocellular system. The system we are investigating consists of only three molecular components, including lipid, anchored DNA, and anchored ruthenium complexes, where these three subsystems exist in a particular environment that provides resource molecules and free energy, where each molecular component is chosen to simplify the overall system [Rasmussen et al., (2003), Rasmussen et al., (2016), Thomsen et al., (2025)]. For example, DNA modified with a lipophilic anchor is used to attach the DNA to the membrane. This removes the significant problem of encapsulation efficiency and removes the need to optimize for the retention of encapsulated DNA after division. The replication of DNA exploits LIDA (Lesion-Induced DNA Amplification), a mechanism that is isothermal and can be tuned over a broad range of temperatures to be compatible with integrated chemistry. The protocell relies on light for energy that requires fewer input components and a process that produces fewer byproducts, thus decreasing the likelihood of unwanted product interactions. The metabolism relies on a ruthenium-based energy transducer. The metabolism produces new lipids that self-assemble into the vesicle membrane as well as ensures the ligation in the LIDA process. The flow of electrons in our metabolic system is unidirectional, thereby reducing unproductive electron transfer events. Note that DNA does not encode RNA or protein in this system, so the protocell does not exploit translation, a highly complex, energetically costly process that is not amenable, thus far, to replication from a few simple component parts. See Fig. 3.1.

More specifically, the protocell consists of a lipid vesicle with a ruthenium complex and DNA anchored to the surface. When exposed to light, the ruthenium in the complex (e.g. ruthenium tris(bipyridine) $[Ru(bpy)_3]^{2+}$) is oxidized to Ru (III), whereby it donates an electron to the bispyridyl ligand. Ru (III) is then re-oxidized by accepting an electron from the guanosine residue within the DNA duplex closest to the ruthenium complex on the lipid vesicle surface. The generated 'electron hole' then propagates up the proximal DNA duplex until it reaches an oxoguanine residue, which is even more easily oxidized than guanine. This process can be described as DNA hole transport. The fuel molecule dihydrophenylglycine acts as a "H source" regenerating the 8-oxoguanine by donating an electron and a proton. Thus, the DNA provides a resulting path for a flux of electrons from dihydrophenylglycine to the ruthenium complex with a speed and efficiency dependent on the sequence of the DNA. The electron that originated from the ruthenium center Ru(II) - the loss of which formed the hole - and which was stored on the bipyridine moiety in the excited state, is donated to a picolinium ester of a fatty acid moiety, resulting in ester cleavage. This generates a deprotected fatty acid/lipid moiety, that is available as building block to be incorporated into the vesicle. The entire process occurs in iterative cycles, because of the fuel molecule dihydrophenylglycine delivering the electron needed to alleviate the electron hole at the 8-oxoguanine site caused by photo-induced oxidation of Ru(II) to Ru(III). Taken together, it is the energy from light that drives further vesicle growth until it reaches a point of unstable morphology and division. See Fig. 3.2.

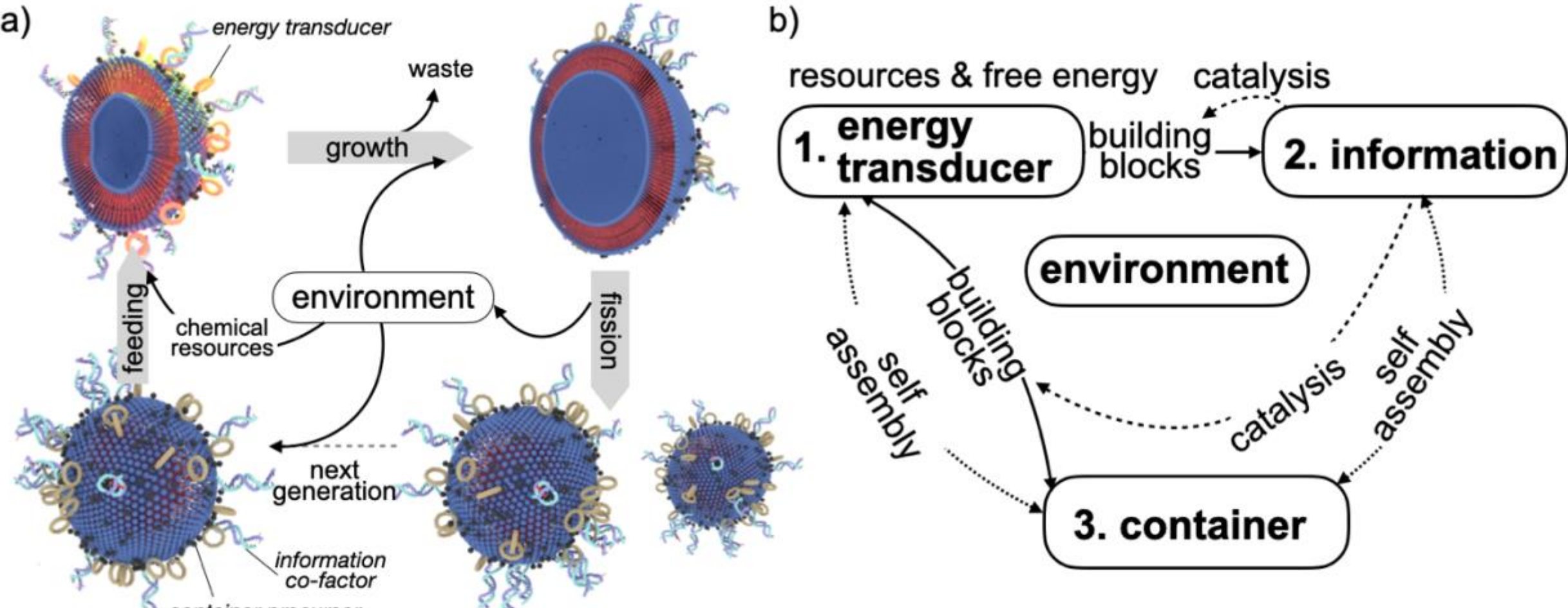

***Figure 3.1*** *(a) Protocellular lifecycle of the system: energy transducer, information, and container in a specially controlled environment that provides resources and free energy. This design has organizational closure and autocatalytic component integration. The protocells are vesicle surface-functionalized with two coupled components: energy transducer and information co-factor. (b) The energy transducer uses external free energy (light) to drive the metabolism and turn resources into building blocks (solid arrows), while the information co-factor controls the metabolic efficiency (dashed arrows) and both are kept together by the container (dotted arrows). The combined energy transducer and co-factor define the metabolism. For details, see text and Fig 3.2.*

These processes are coupled with replication of the DNA on the vesicle surface to ensure that each daughter vesicle receives DNA from the parent vesicle. This DNA replication is based on the LIDA process, in which fragments are ligated at the template, where defined and appropriately placed lesions enable product dissociation, generating new templates for the next round of fragment ligation. Therefore, feeding smaller resource fragments that are complementary to the duplex DNA allows the information carrier at the vesicle surface to be replicated. To couple this replication process with the light-driven vesicle growth, one of the resource fragments contains a protecting group, which is removable under oxidative conditions. Upon binding of the fragment to the DNA template and forming a nicked duplex, this protecting group donates an electron to the electron hole induced by light-driven Ru oxidation. Consequently, the protecting group is released through oxidative removal, allowing the two smaller resource oligomer fragments to be ligated into a new strand. Thus, both container growth and information replication depend on the light-driven oxidation of the ruthenium complex, thereby coupling the two processes.

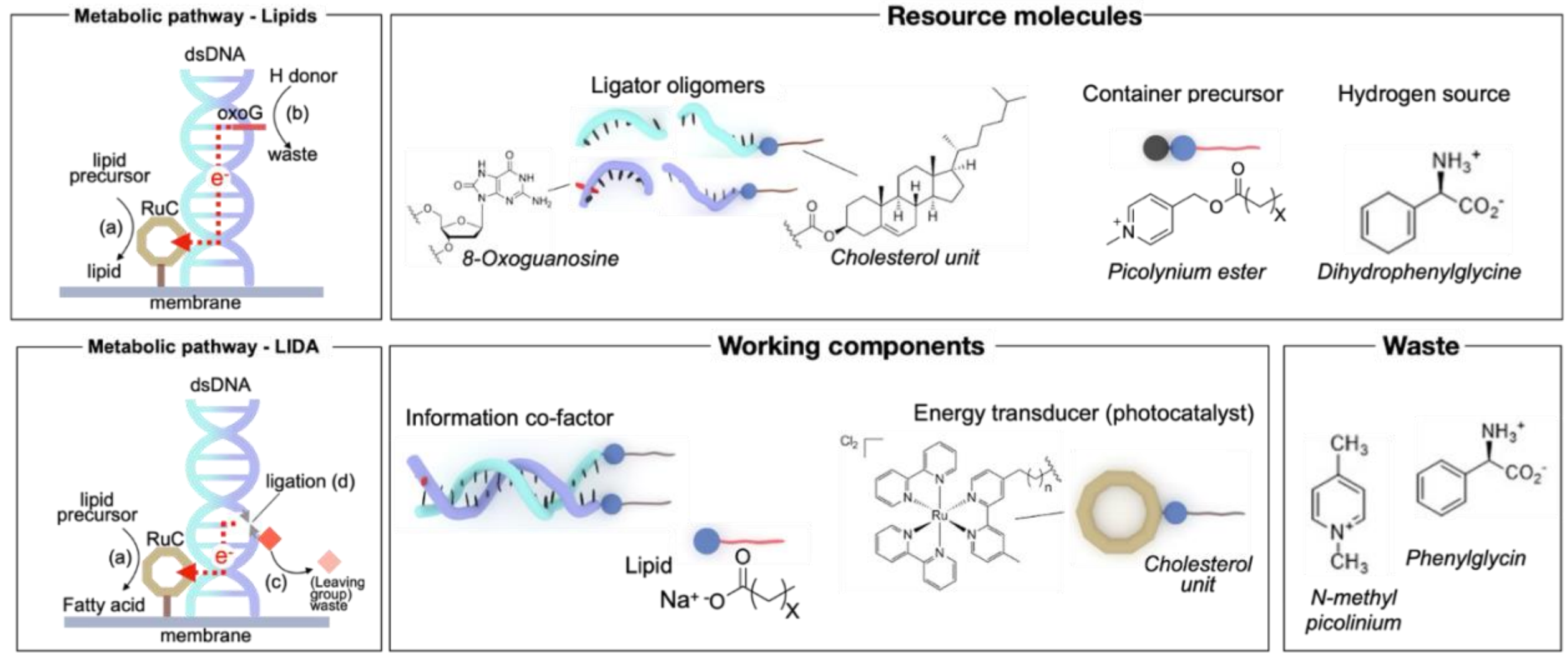

***Figure 3.2*** *Top: Protocellular processes and molecular components. Panels (a) and (b) Resulting electron path for DNA ligation and fatty acid production respectively. (c) Resource molecules; (d) Building blocks for anchored DNA co-factor and container fatty acids; (e) Anchored ruthenium complex-based energy transducer; (f) Waste molecules. See text for details.*

# 4 Protocellular energetics

To estimate the free energy changes associated with each lifecycle in the presented protocell, we first estimate the free energy input and transformations associated with the metabolic building block generating processes. These resulting building blocks are new fatty acids for the container growth and new DNA co-factor molecules that modulates the metabolism. However, to obtain a more complete picture of the free energy changes associated with a protocellular lifecycle, we also analyze the free energies associated with the many important self-assembly processes. These include vesicle assembly, anchoring of the energy-transducer and co-factor to the vesicle membrane, hybridization and de-hybridization processes associated with the co-factor replication. Finally, we also estimate the membrane bending energies associated with vesicle growth, deformation, and reproduction.

## 4.1 Energetics of fatty acid production

This metabolic reaction network is primarily driven by photo-energy combined with reaction steps driven by free energy from energy rich molecules. This network generates fatty acid molecules that self-assembles into vesicles. The complete reaction network is rather complicated, see Fig 4.1.1, but we present it up front to provide an overview of the different parts of the reaction network here indicated in different colors. The free energy changes indicated in Fig. 4.1.1 are either from the literature or derived in later in this section 4.1. We will present and analyze each part of the network separately, starting with the overall resulting fatty acid production around Fig 4.1.2. For more details, please see DeClue et al. (2009), Bornebusch et al. (2021), and Thomsen et al. (2025).

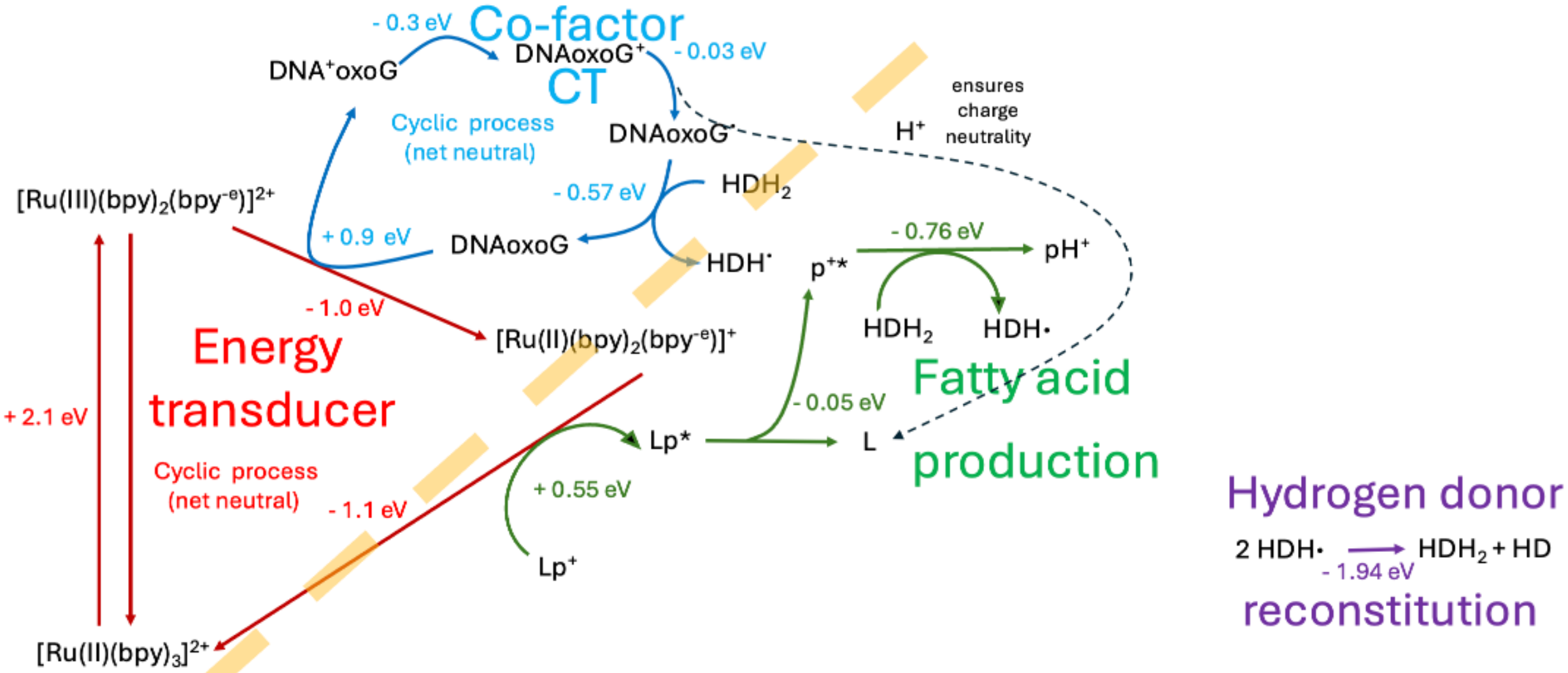

***Figure 4.1.1*** *Summary of main reaction network for fatty acid production. Photo-driven reactions for the ruthenium complex energy transducer are indicated in red, while the reactions for the DNA co-factor charge transfer (CT) are indicated in blue. We will refer to these reactions above the manila dotted line as the 'upstream' reactions. The combined energy transducer and co-factor operate as a catalyst for the fatty acid production reactions, which are shown in green below the manila dotted line, and this lower part we will refer to as the 'downstream' reactions. Finally, the reconstitution reaction for the hydrogen donor is indicated in purple (far lower right), which also couples the upstream and downstream processes. Detailed discussions of this reaction network are found in [DeClue et al. (2009), Bornebusch et al. (2021), and Thomsen et al. (2025)]. The free energies shown for each of the reactions are either from the literature or derived and discussed here in subsection 4.1. Note the two cyclic processes (red and blue) are energetically net neutral.*

The resulting reaction of the metabolic network presented in Fig. 4.1.1 is shown and discussed in Fig. 4.1.2.

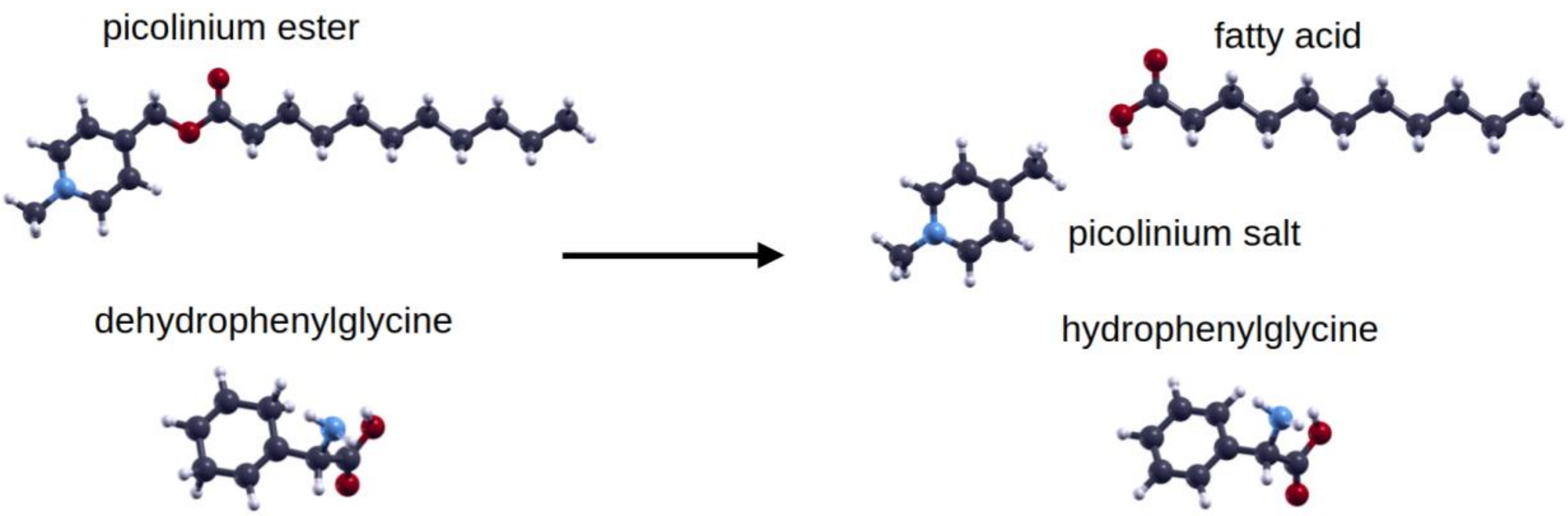

***Figure 4.1.2*** *Overall resulting reaction omitting the reactions involving the combined energy transducer and co-factor as the two together act as a catalyst for the reaction. The picolinium ester is deprotected to form decanoic (fatty) acid and picolinium salt, while the dihydrophenylglycine hydrogen donor delivers two hydrogen atoms, one to reconstitute the oxoguanine and one to the picolil radical to form the stable picolinium salt.*

As indicated in Figs 4.1.1 the production of fatty acid involves several 'upstream' reactions, starting with photochemistry, then charge transfer between energy-transducer (RuC) and co-factor (dsDNA) as well as charge transport within co-factor, which we present and discuss below.

The photo-activation energy is $\Delta G$ = 2.1 eV or 202.27 kJ/mol for each produced fatty acid molecule, and the quantum yield is estimated to be 0.0044 or 0.44% of the actual (free) photo energy input per metabolic step [DeClue et al. (2009)]. Thus, by far the most photons do not result in an excitation of the ruthenium complex.

The first part of the reaction involves photo-activation of the ruthenium complex and if an electron from the DNA cofactor can be delivered to the ruthenium such that Ru(III) + -e -> Ru(II), before the back reaction can occur, then the lipid producing reactions can proceeded from $[Ru(II)(bpy)_2(bpy^{-e})]^+$. A back reaction occurs when the photo activated electron jumps back from one of the dipyridine rings onto the ruthenium atom from where it originated. See Fig. 4.1.1.

In the reaction Ru(III)C + -e -> Ru(II)C, an electron is captured from a guanine within a dsDNA at the site where the ruthenium complex, RuC, is intercalated. The energy needed to extract an electron from a guanine molecule is about +0.9 eV [Psciuk et al. (2012)] and thus less than the available -1.0 eV. This process leaves an 'electron hole' – a positive charge – within the dsDNA. This hole now diffuses within the dsDNA until it reaches an oxoguanine at one of the two strands. The oxoguanine then delivers an electron that fills the hole, but as a result it successively also loses a hydrogen atom and then becomes a radical. We will refer to these processes as the 'upstream' reactions for the decanoic acid production as they that involve the initial photoactivation followed by the DNA charge transfer, oxo-guanine electron donation and proton loss, succeeded by oxo-guanine reconstitution by means of a hydrogen donor.

Writing these reactions, we may add inert components to the reactions on both side of the reaction arrow, which will be useful when we below are ready to also consider the 'downstream reaction' in the decanoic acid production, which are the reaction that directly involves the transformation of the picolinium ester into decanoic acid.

To each set of moieties in each state $S_i$ (reactant or product set) we can assign a total free energy expressed by $G_i$, i = 1, … , 11. As will become apparent later, we use these reactions to estimate the free energy change for every process in the metabolic production of fatty acids. The fragments that are unchanged with respect to the previous state are indicated in grey.

light activation
of 2 $Ru(II)C]^{2+}$
2 x 2.1 eV (added free energy from light)

$S_1$ = 2 $[Ru(II)C]^{2+}$(dsDNAoxoG) +3 $HDH_2$ + 2 $Lp^+$ →

(1a)

2 -e donated from G to 2 R(III)
2 x 1.0 eV followed by DNA hole diffusion (spontaneous)

$S_2$ =2 $[Ru(III)C^{-e}]^{2+}$(dsDNAoxoG) +3 $HDH_2$ + 2 $Lp^+$ →

(1b)

2 oxoG donation of 2 -e (spontaneous)

$S_3$ =2 $[Ru(II)C^{-e}]^+$($dsDNA^+$oxoG) +3 $HDH_2$ + 2 $Lp^+$ →

(1c)

2 $H^+$ lost from 2 $oxoG^+$ (spontaneous)

$S_4 = 2\ [Ru(II)C^{-e}]^+(dsDNAoxoG^+) + 3\ HDH_2 + 2\ Lp^+ \rightarrow$

(1c)

2 reconstitutions of 2 oxoG by 2 $HDH_2$

$S_5 = 2\ [Ru(II)C^{-e}]^+(dsDNAoxoG^*) + 2\ H^+ + 3\ HDH_2 + 2\ Lp^+ \rightarrow$

(1d)

reconstitution of $HDH_2$
and HD waste production
(externally added, originally from $HDH_2$)

$S_6 = 2\ [Ru(II)C^{-e}]^+(dsDNAoxoG) + 2\ H^+ + 2\ HDH^* + 1\ HDH_2 + 2\ Lp^+ \rightarrow$

(1e)

$S_7 = 2\ [Ru(II)C^{-e}]^+(dsDNAoxoG) + 2\ H^+ + 2\ HDH_2 + 1\ HD + 2\ Lp^+$

We now switch the focus to the 'downstream' reactions for the decanoic acid production:

2 x 1.1 eV (added free energy, originally from light)
2 x activated -e from
2 $[Ru(II)C^{-e}]^+$

$S_7 = 2\ [Ru(II)C^{-e}]^+ dsDNAoxoG) + 2\ H^+ + 2\ HDH_2 + 1\ HD + 2\ Lp^+ \rightarrow$

(2a)

$S_8 = 2\ [Ru(II)C]^{2+}(dsDNAoxoG) + 2\ H^+ + 2\ HDH_2 + 1\ HD + 2\ Lp^{+*} \rightarrow$

(2b)

reconstitution of picolinium salt
and formation of fatty acid
(by equilibrium of $H^+$ with environment)

$S_9 = 2\ [Ru(II)C]^{2+}(dsDNAoxoG) + 2\ HDH_2 + 1\ HD + 2\ L + 2\ p^{+*} \rightarrow$

(2c)

(externally added, originally from $HDH_2$)
reconstitution of $HDH_2$
and HD waste production

$S_{10} = 2\ [Ru(II)C]^{2+}(dsDNAoxoG) + 2\ HDH^* + 1\ HD + 2\ L + 2\ pH^+ \rightarrow$

(2d)

$S_{11} = 2\ [Ru(II)C]^{2+}(dsDNAoxoG) + HDH_2 + 2\ DH + 2\ L + 2\ pH^+.$

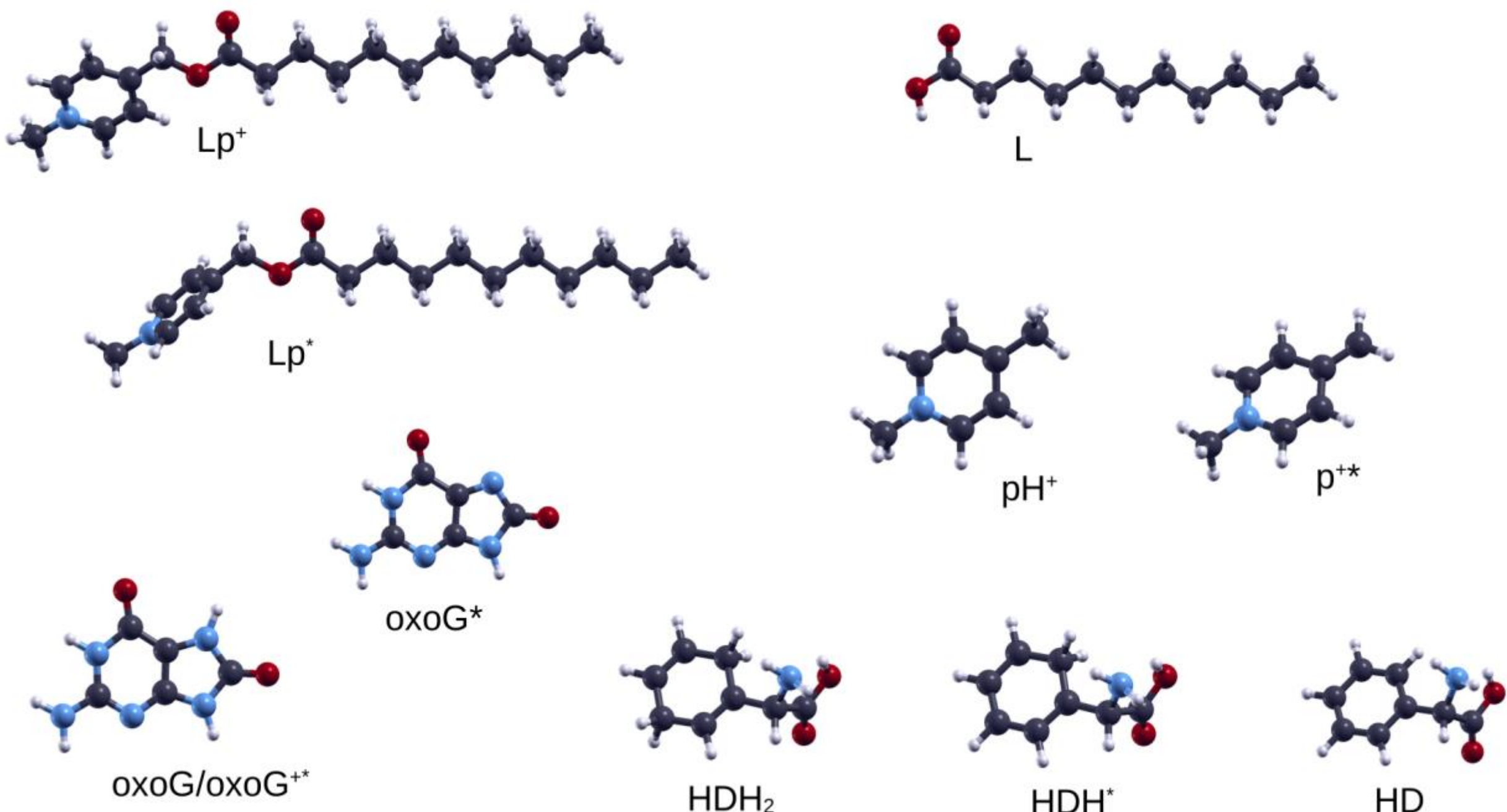

***Figure 4.1.3*** *Molecular structures for which the free energies were calculated using density function theory (DFT) as described in SI.1. This information, together with other known reaction energies, enable us to write a closed set of equations, from which the free energy change of the remaining metabolic reactions can be deduced.*

Counting the free energies known from the literature as well as the ones calculated here using DFT for some of the molecules, see Fig 4.1.3, we can write a set of 29 coupled linear equations and for the remaining unknowns. This equation system is shown and described in SI.2. From the solutions to the SI.2 equation system we can obtain the free energy changes along the reaction path expressed above in Eq. (1a) → (2d), shown in Fig. 4.1.4.

For the calculations in Figs. 4.1.4 – 4.1.6, we have set the light excitation to be + 2.1 eV, the Ru electron capture to – 1.0 eV, the hole diffusion to – 0.3 eV, the oxidation of oxoG to + 0.6 eV, and the deprotonation of oxoG to **–** 0.03 eV, see SI.2 for details.

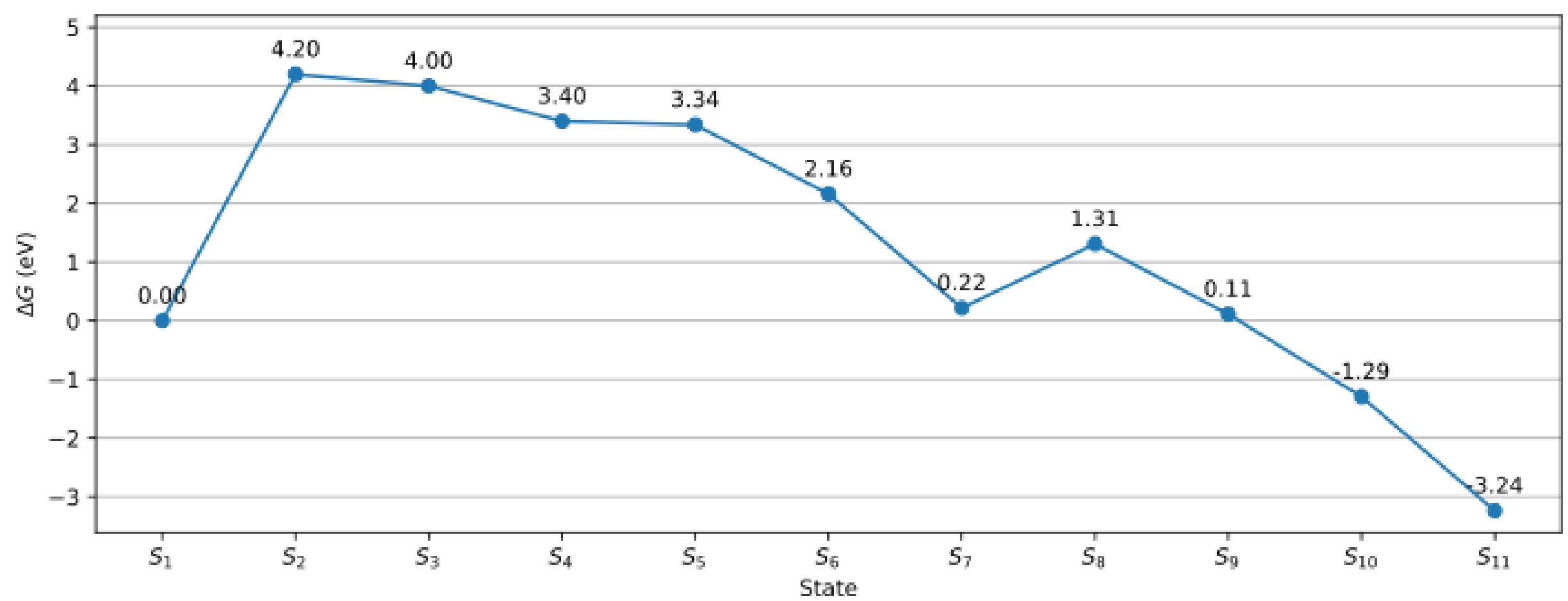

***Figure 4.1.4*** *Free energy for the reaction steps involved in the production of decanoic acid. Values in eV given for each state $S_i$ where $S_1$ is set to 0, and approximations for the DFT input are based on B3LYP. See text and SI2 for details. Recall from reactions 1a – 2d that the depicted processes involve two photoexcitations and result in the production of two decanoic acid molecules. The hydrogen donation to the picolil radical is not spontaneous ($S_7 \rightarrow S_8$, $\Delta G > 0$), but it happens because of the reaction couplings in the system, as the full reaction system is verified experimentally [DeClue et al. (2009)].*

From the solutions to the SI.2 equation system we can also depict the 'upstream' and 'downstream' processes as found in Figs. 4.1.5 and 4.1.6.

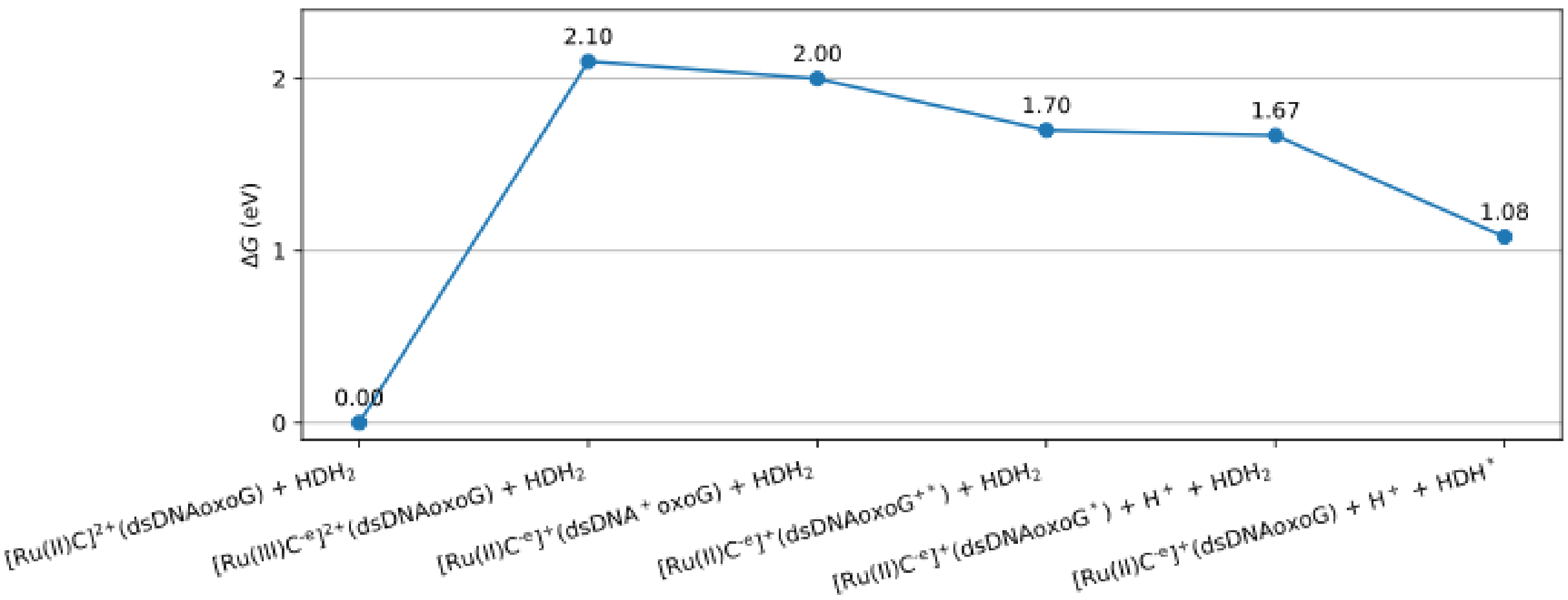

***Figure 4.1.5*** *Free energy of the upstream part of the reaction system. Values in eV given for each state $S_i$ where $S_1$ is set to 0, where the DFT input are based on B3LYP. See text and SI2 for details. Initial photoexcitation followed by charge transfer in the dsDNA co-factor. Electron donation to the ruthenium complex (RuC) is originally from a guanine (G) within the dsDNA next to the site where the RuC is intercalated. This is followed by hole diffusion until it meets the oxoG site from which an electron is delivered to the hole whereafter the oxoG loses a $H^+$ atom. Oxoguanine is eventually repaired by a sacrificial hydrogen donor, dihyrdophenylglycine. See text for details.*

Thus, the overall free energy change from input materials (picolinium ester and dihydrophenyleglycine) to output materials (decanoic acid) in the coupled processes is - 3.24 eV/2 = - 1.62 eV, recall the discussion around Fig. 4.1.3, while the free energy input for the photoactivation process is + 2.1 eV. As pointed out previously, the combined energy transducer (RuC) and co-factor (dsDNAoxoG) complex acts as catalyst for the fatty acid production, the photoactivation may be viewed as free energy needed to overcome a reaction barrier.

In SI2 different

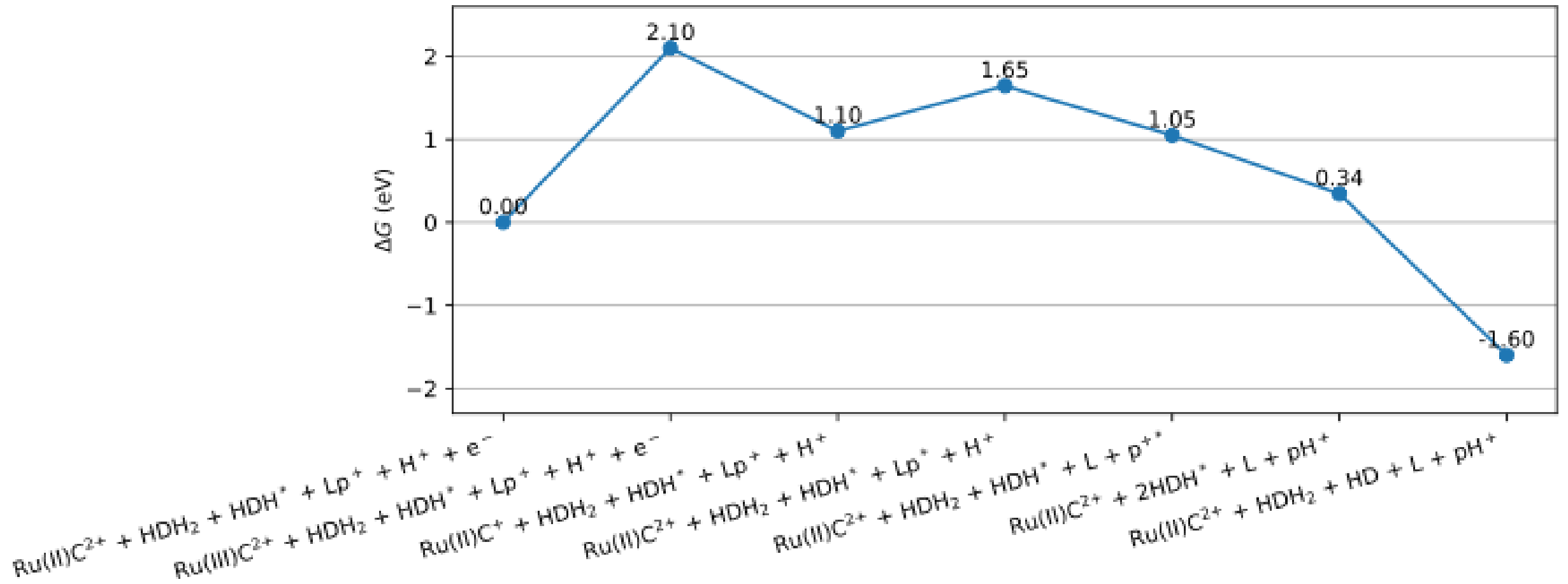

**Figure 4.1.6** *Free energy in downstream part of the fatty acid production where the DFT input are based on B3LYP approximations and the initial state is set to 0. See SI.2 for details. After the upstream processes, an energy rich electron is delivered from the RuC to the picolinium ester, which is in the process step 3 to 4 in the figure. This results in a deprotection of the picolinium ester, formation of fatty acid and the reconstitution of the picolyl radical by dehydrophenylglycene. See text for details.*

## 4.2 Information (co-factor) replication energetics

DNA replication involves several hybridization and de-hybridization processes as well as a non-trivial, two-step oligomer-oligomer ligation process, which are discussed in the following. The DNA co-factor replication is based on a lesion induced DNA amplification (LIDA) process that circumvents product inhibition, which is coupled to the metabolism through a deprotection and a ligation step. For details on the LIDA process and its kinetics, e.g. see Alladin-Mustan et al. (2015) and Park et al. (2023), or Thomsen et al. (2025) in the context of protocellular information replication.

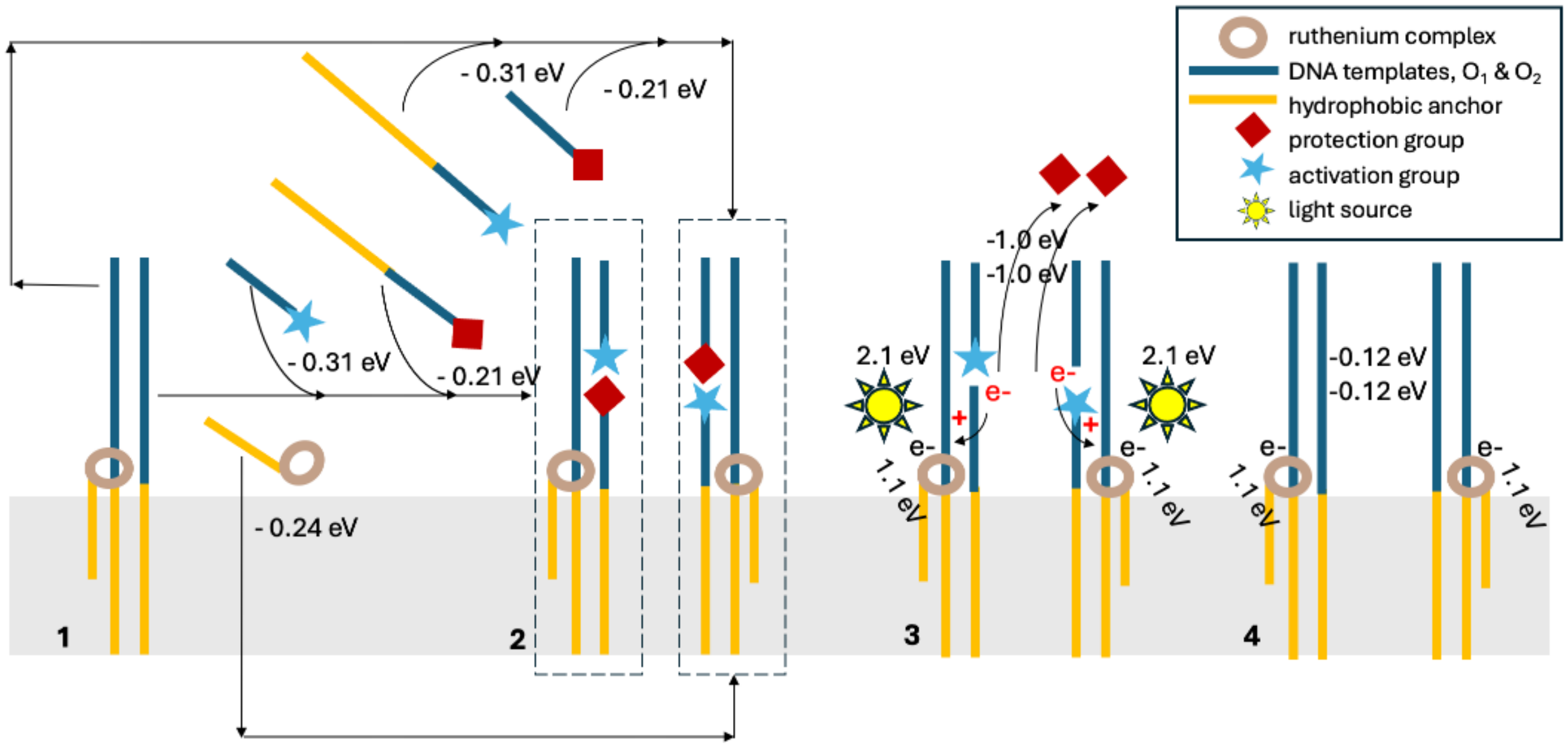

**Figure 4.2.1**: *Information co-factor replication at the vesicle surface based on the lesion induce DNA amplification (LIDA) process. The initial state is indicated by (1). Next resource molecules for the DNA co-factor replication are added (1) -> (2): a ruthenium complex with an anchor (grey ring) and four oligonucleotides (blue), two with anchors, either with a leaving group (red diamond) or with an activation group (blue star) at the ligation site. (2) These resource molecules self-assemble at the vesicle surface and/or through templating or intercalation. Note that the vesicle anchoring processes are examined separately in Section 4.3. (3) Upon light irradiation the RuC is activated that results in hole formation in the dsDNA (recall Section 4.1). The leaving group (red diamond) is oxidized once the hole reaches the ligation site and leaves (deprotects) the oligomer. (4) The following ligation is driven by the activation group (blue star). Note that (4) also depicts the energy rich electron that is successively used for fatty acid production. See text for details.*

In the following we have used SantaLucia & Hicks (2004) as well as NUPACK to estimate the free energies of hybridization.

The hybridization energies for 9-base oligonucleotides $O_1$ and $O_2$ utilized in the LIDA process are typically between -16 kJ/mol and - 35 kJ/mol, while the hybridization energy for an 18 bp dsDNA may be between -40 and -60 kJ/mol. Note that these sequences both contain a mismatch as well as an abase or a bulge at the ligation site.

As an example, let us assume $\Delta G_{dsDNA}$ = 54 kJ/mol or 0.56 eV for each molecule.

Thus, we may assume:

$\Delta G_{i\text{-}o1}$ = - 20 kJ/mol or - 0.21 eV for each oligomer$_1$-template molecule (with or without anchor),
$\Delta G_{i\text{-}o2}$ = - 30 kJ/mol or - 0.31 eV for each oligomer$_2$-template molecule (with or without anchor),
$\Delta G_{i\text{-}nick}$ = - 52 kJ/mol or - 0.54 eV for each nicked duplex molecule, so
$\Delta G_{ds}$ = - 2 kJ/mol or - 0.02 eV hybridization energy gain for each molecule after ligation.

Both the dsDNA, the ssDNA and two of the oligonucleotides are anchored to the vesicle surface. The free energies associated with these self-assembly anchoring processes are discussed separately in subsection 4.3 together with the anchoring of the ruthenium complex.

Once a (partial) DNA duplex is formed the composed of the anchored ssDNA template and the anchored oligonucleotide $O_1$ a nearby anchored ruthenium complex can intercalate. As the observed binding constant for this process is reported to be about $K_b = 10^{-4}$/M [Xu et al (2005), Wu et al. (2005)], the free energy of intercalation can be calculated as

$$\Delta G = RT \ln (K_b) = - 22.8 \text{ kJ/mol or } - 0.24 \text{ eV/molecule.}$$

The oligomer-oligomer ligation requires a deprotection of one of the oligomers. We assume the oligomer deprotection occurs by electron delivery from an oxidative cleavage protection group OCG at the sugar moiety into an electron hole within the dsDNA that releases the cleaved group OCG*. As this reaction step is not yet realized in the lab, the energetics of the deprotection is just an assumption for the following estimations. The electron hole in the dsDNA is created by the free energy input of 1.0 eV from the photo activation process as discussed in subsection 4.1. For simplicity, we further assume that any potential downstream reactions with the free OCG* molecule do not interfere with the protocellular processes, so that we can ignore them in our further estimations.

After deprotection, the ligation process that creates a phosphodiester (or phosphoamidate) bond requires activation that can be provided by a phosphate imidazole (IMZ). The standard free energy released for hydrolysis of UMP-imidazole is $\Delta G = - 32.0$ kJ/mol [Maguire et al (2021)] or - 0.33 eV/molecule, while the standard free energy change for the hydrolysis of a DNA phosphodiester bond is $\Delta G = -5.3$ kcal/mol (- 22.2 kJ/mol) or - 0.23 eV/molecule [Dickson et al. (2000)]. Thus, a resulting free energy change of about -0.1 eV.

The processes associated with each of the two oligomer ligations in the dsDNA co-factor replication process can be expressed as follows:

$[Ru(II)C]^{2+}$(ssDNA)($O_1$ OCG)(IMZ_$O_2$) →
+ 2.1 eV, [Vlcek et al. (1995)]

$[Ru(III)C^{-e}]^{2+}$(ssDNA)($O_1$_OCG)(IMZ_$O_2$) →
- 0.1 eV, recall Sec. 4.1, Fig. 4.1.5.)

$[Ru(II)C^{-e}]^{+}$(ssDNA_$O_1^{+}$_OCG)(IMZ_$O_2$) →
$- 0.9 \text{ eV} \leq \Delta G < 0$, assumption)

$[Ru(II)C^{-e}]^{+}$(ssDNA_$O_1$_) (IMZ_$O_2$) + OCG* →
- 0.33 eV + 0.23 eV = - 0.1 eV due to activation and ligation [(Maguire et al. (2021) and Dickson et al. (2000)] as well as - 0.02 eV in hybridization energy due to ligation, resulting in - 0.12 eV.

$[Ru(II)C^{-e}]^{+}$(dsDNA) + OCG* + IMZ.

A summary of the free energy changes for the DNA co-factor replication processes is depicted in Fig. 4.2.2. Note that Fig. 4.2.2(4) also depicts the energy rich electron of + 1.1 eV that is successively used for fatty acid production in the 'downstream' part of this process, recall Section 4.1.

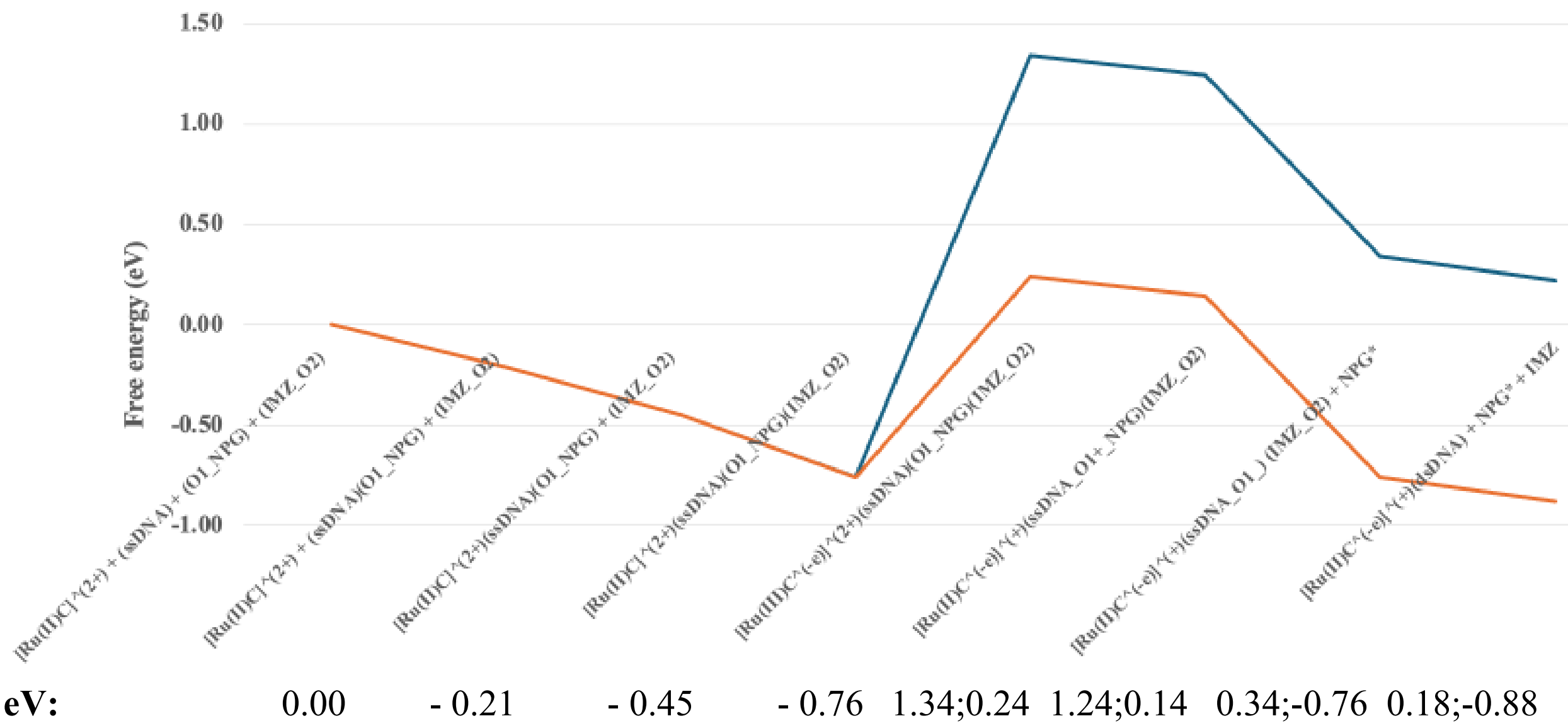

**Figure 4.2.2** *Energetics of the DNA co-factor replication related processes with relative free energy change indicated under the graph for each reaction compared to the initial state that is set to 0. Note that this graph depicts the replication of a single DNA template, while the co-factor replication requires replication of two templates (plus and minus). Initially, template-oligomer hybridization and intercalation processes are depicted, followed by photoactivation, deprotection and ligation. Blue curve includes the energy rich electron at one of the bipyridine rings, while the orange curve only includes the part of the photo-induced free energy that is directly utilized in the DNA replication process, which doesn't include the energy rich electron. See text for more details.*

The external free energy input for each dsDNA co-factor replication is thus two ruthenium complex excitations plus the energetics of the two activation groups 2 x (2.1 eV + 0.33 eV) = 4.88 eV, while the hybridization energy of a new dsDNA (co-factor) is about - 0.56 eV per new DNA duplex and – 0.24 eV for the new RuC intercalation, which are both self-assembly processes. In the following free energy estimations these hybridization and intercalation processes are included, while the vesicle anchoring processes of the RuC and dsDNA are examined separately in Section 4.3.

Thus, the free energy difference between the initial and the end state of the dsDNA co-factor replication process is:

initial oligomer hybridizations (- 0.52 eV) + one new RuC intercalation (- 0.24 eV)
+ two activations 2 x (- 0.33 eV) + (part of) two photoactivation 2 x (+ 1.0 eV)
+ two hole formations and deprotections 2 x (- 0.1 eV + - 0.9 eV)
+ two ligations 2 x (+ 0.23 eV) + hybridization correction due to ligation (- 0.02eV) = - 0.98 eV.

Note that the free energy estimation for the DNA replication process include covalent bond breakage and bond formation as well as self-assembly processes, but not the self-assembly processes associated with vesicle anchoring of the dsDNA and the RuC.

## 4.3 Container energetics

For simplicity we assume the start situation is characterized by one vesicle, and the end situation is characterized by two vesicles, both identical to the original. Thus, the free energy change only associated to the vesicle change is the addition of one identical vesicle, as metabolic transformation free energies are calculated elsewhere. Also, the free energy changes due to feeding of container precursors (picolinium ester) are implicitly accounted for as these precursors are eventually transformed into container building blocks.

### 4.3.1 Vesicle self-assembly

To estimate the free energy associated with the container self-assembly we need to know the vesicle size. Assuming a vesicle of diameter D = 1 μm = 1 x $10^{-6}$ m and radius r = 0.50 x $10^{-6}$ m, the vesicle outer surface area is thus A = 4 π $r^2$ = 4 π (5.0 x $10^{-7}$)$^2$ $m^2$ = 3.14 x $10^{-12}$ $m^2$, and the area of a single fatty acid lipid head [Danov et al. (2006)] is 0.22 $nm^2$ = 0.22 x $10^{-9}$ x $10^{-9}$ $m^2$ = 0.22 x $10^{-18}$ $m^2$. Thus, the number of lipid molecules in the vesicle bilayer is $N_b$ = 2 x (3.14 x $10^{-12}$ $m^2$) / (0.22 x $10^{-18}$ $m^2$) = 2.9 x $10^7$ ~3 x $10^7$.

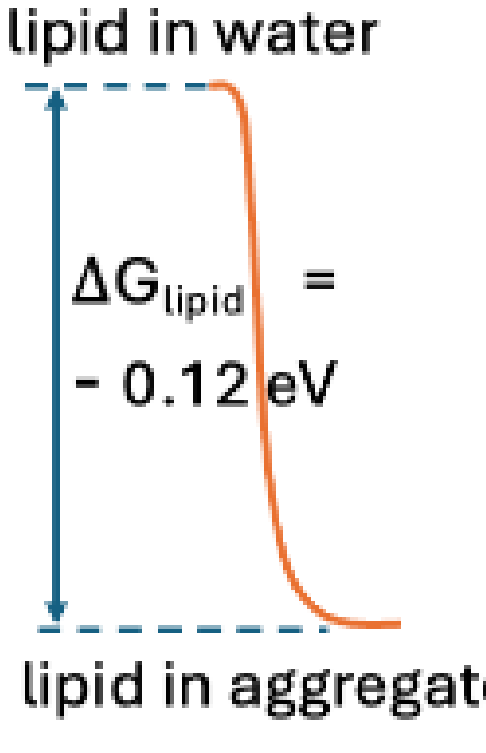

**Figure 4.3** *Lipid self-assembly. Each of the about 3x$10^7$ fatty acids that join a vesicle lowers the free energy of the system. Each additional fatty acid originally joins the vesicle as a picolinium ester resource that is converted into a fatty acid, so the final free energy contribution from the self-assembly process is that of a fatty acid joining the vesicle from the solution.*

For each lifecycle in principle one new identical vesicle is generated, which presents a free energy addition per mol of fatty acid due to self-assembly. As the standard Gibbs free energy of formation of decanoate is estimated in Section 4.1 for the single-molecule (unimer) state and not for the vesicle membrane state, we can compute the standard Gibbs free energy difference between the unimer state and the vesicle membrane state.

This means that we need to calculate the difference in standard Gibbs free energy between decanoate in the unimer state and in the vesicle membrane state.

Decanoate synthesized by the combined information co-factor and photocatalyst initially exists at the interface of the vesicle surface and the aqueous phase where its dissociation state is determined by the pH in the external solution.

As an unimer in the aqueous phase, at equilibrium, the chemical potentials of decanoic acid (DA), $\mu_{\mathrm{uni}}^{(\mathrm{DA})}$, decanoate ($D^-$), $\mu_{\mathrm{uni}}^{(\mathrm{D}^-)}$, and proton ($H^+$), $\mu_{w}^{(\mathrm{H}^+)}$, satisfy the relationship

$$\mu_{\mathrm{uni}}^{(\mathrm{DA})} = \mu_{\mathrm{uni}}^{(\mathrm{D}^-)} + \mu_{w}^{(\mathrm{H}^+)}.$$

The chemical potentials of DA and $D^-$ in the unimer state, $\mu_{\mathrm{uni}}^{(\mathrm{DA})}$and $\mu_{\mathrm{uni}}^{(D^-)}$, are expressed in terms of their standard chemical potentials, $\mu_{\mathrm{uni}}^{0\,(\mathrm{DA})}$and $\mu_{\mathrm{uni}}^{0\,(\mathrm{D}^-)}$, and their molar concentrations, $c_{\mathrm{uni}}^{(\mathrm{DA})}$ and $c_{\mathrm{uni}}^{(\mathrm{D}^-)}$, respectively, as

$$\mu_{\mathrm{uni}}^{(DA)} = \mu_{\mathrm{uni}}^{0\,(DA)} + k_B T \ln\left(c_{\mathrm{uni}}^{(DA)}/c^0\right)$$

$$\mu_{\mathrm{uni}}^{(D^-)} = \mu_{\mathrm{uni}}^{0\,(D^-)} + k_B T \ln\left(c_{\mathrm{uni}}^{(D^-)}/c^0\right) - e\psi_{\mathrm{w}},$$

where the standard concentration $c^0 = 1\mathrm{M}$. Here, $e\psi_{\mathrm{w}}$ is the electrostatic potential energy of the ionic molecule, where $\psi_{\mathrm{w}}$ is the electrostatic potential in water phase and $e$ is the elementary charge ($1.602\times 10^{-19}$ C).

The chemical potential of the proton is expressed in terms of its molar concentrations, $c_{w}^{(H^+)}$, as

$$\mu_w^{(H^+)} = k_B T \ln\left(c_w^{(H^+)}/c^0\right) + e\psi_w,$$

where the standard chemical potential of protons is set as $\mu_{H^+}^0 = 0$.

We next evaluate the chemical potentials of DA and D⁻ within the vesicle membrane.
Let the number of DA molecules in a vesicle be $M$, with a standard chemical potential $\mu_{\mathrm{ves}}^{0\,(DA)}$, the number of D⁻ molecules be $N$, with a standard chemical potential $\mu_{\mathrm{ves}}^{0\,(D^-)}$, and the molar concentration of vesicles in solution be $c_{\mathrm{ves}}$. Then, the chemical potential of one vesicle is expressed as

$$\mu_{\mathrm{ves}} = M\mu_{\mathrm{ves}}^{0\,(DA)} + N\mu_{\mathrm{ves}}^{0\,(\mathrm{D}^-)} + k_B T\left[M \ln x_{\mathrm{ves}}^{DA} + N \ln x_{\mathrm{ves}}^{\mathrm{D}^-}\right] + k_B T \ln(c_{\mathrm{ves}}/c^0) - Ne\psi_{\mathrm{ves}}$$

where $x_{\mathrm{ves}}^{DA} = M/(M+N)$ is the mole fraction of DA in the membrane, $x_{\mathrm{ves}}^{\mathrm{D}^-} = N/(M+N)$ is that of D⁻, $\psi_{\mathrm{ves}}$ is the electrostatic potential at vesicle membrane. The first and second terms represent the standard chemical potentials of DA and D⁻ in the vesicle membrane, the third term accounts for the entropy of mixing, the fourth term corresponds to the translational entropy of vesicles, and the fifth term is the electrostatic potential contribution.

Elaborating on the equations above, it can be derived that the resulting free energy difference between decanoate in solution and in the vesicle is

$$\mu_{\mathrm{uni}}^{0\,(\mathrm{D}^-)} - \mu_{\mathrm{ves}}^{0\,(\mathrm{D}^-)} = \Delta G_f^0[D^-{}_{\mathrm{uni}}] - \Delta G_f^0[D^-{}_{\mathrm{ves}}] = -\,k_B T \ln\left(\frac{\mathrm{CVC}(D^-)}{5}\right) = 4.5 k_B T \approx 0.12 \text{ eV per molecule.}$$

thus favoring the vesicle phase. The detailed derivation for this is presented and discussed in SI.3 (Appendix), using $pK_a = 4.9$, assuming pH = 8, CVC = 57 mM for decanoate ($D^-$), and N/M ~ 1 in the membrane.

For the above estimation we could alternatively have used – 2.3 kJ/mol per $CH_2$ group as reported by Budin et al. (2014). This corresponds to $-6.242\mathrm{x}10^{18}$ eV/J x $2.3\mathrm{x}10^3$ J/mol = $-14.36\mathrm{x}10^{21}$ eV/mol or ( $-1.436\mathrm{x}10^{22}$ eV/mol)/($6.022\mathrm{x}10^{23}$/mol) = –0.024 eV per $CH_2$ group and thus –0.21 eV for DA. As expected, this value is higher value than our above estimates as we also include the weakening impact of the electrostatic headgroup interactions.

The total free self-assembly energy from a newly added vesicle is thus $-0.12$ eV/decanoate $2.9 \times 10^7$ decanoate/vesicle = $-3.48 \times 10^6$ eV/vesicle.

### 4.3.2 Vesicle fission

Next, we evaluate the difference in elastic energy for the simplest case where a single vesicle divides into two identical daughter vesicles where each is identical to the original mother vesicle.

Since decanoic acid/decanoate molecules undergo flip-flop motion between the two layers within the membrane, the elastic energy is described by the spontaneous curvature model [Seifert et al., (1991), Urakami et al. (2021)].

According to the spontaneous curvature model, the membrane elastic energy is given by

$$F_{\mathrm{sp}} = \frac{\kappa}{2}\oint dA\,(C_1 + C_2 - C_0)^2 + \kappa_G \oint dA\, C_1 C_2$$

where $C_i$ ($i$ = 1,2) are the principal curvatures of the membrane, $C_0$ is the spontaneous curvature originating from bilayer asymmetry, and $\kappa$ and $\kappa_G$ are the bending rigidity and Gaussian curvature rigidity, respectively. The first term corresponds to the bending energy due to vesicle deformation, and the second term represents the Gaussian curvature contribution, which depends on vesicle topology. Here, we adopt the rough estimate $\kappa_G \simeq - \kappa$ [Landau et al. (1986)]. For a spherical vesicle (initial radius $R_0$) to deform into a limiting shape and subsequently divide, spontaneous curvature is required [Lipowsky (2021), Imai et al. (2022)]. In this system, information co-factors and photocatalysts are anchored only on the outer leaflet of the membrane and so is the metabolic production of new decanoic acid/decanoate. This geometric asymmetry is here assumed to generate the spontaneous curvature necessary for deformation into the limiting shape and for division [Steinkühler et al. (2020)].

Theoretically, the spontaneous curvature required for a vesicle to attain the symmetric limiting shape (two equal-sized spheres connected by a very thin neck) is $C_0 R_0 = 2.4$. Therefore, we assume $C_0 R_0 = 2.4$ for the vesicles studied here. Note that in phospholipid vesicles with GFP anchored, $C_0 R_0$ ranges from 2 to 120 [Steinkühler et al., (2020)].

Then, we estimate elastic energy difference between the initial spherical vesicle state and the final two spherical vesicle (two equal-sized vesicles) state, where an initial spherical vesicle has radius $R_0 = 0.5$ μm, $C_1 = C_2 = 1/R_0 = 2\ \mu\text{m}^{-1}$, spontaneous curvature $C_0 = 2.4/R_0$, and bending rigidity $\kappa = 10\ k_B T$. To our knowledge, the bending rigidity of decanoic acid/decanoate vesicle membranes has not been reported; therefore, we used $\kappa = 10\ k_B T$, measured for vesicles prepared from a 1:1 mixture of decanoic acid and decanol [Misuraca et al. (2022)]. Using the surface area $A = 4\pi R_0^2$ and the Gauss–Bonnet theorem ($\oint dA\, C_1 C_2 = 4\pi$ for a sphere), the elastic energy of the initial vesicle is

$$F_{\text{sp}}^{\text{ini}} = (\kappa/2)(1/R_0 + 1/R_0 - 2.4/R_0)^2 4\pi R_0^2 + 4\pi(-\kappa) = -3.68\pi\kappa = -116\ k_B T$$

After reproduction, two spherical vesicles of radius $R_0$ and spontaneous curvature $C_0 = 2.4/R_0$ are formed, yielding

$$F_{\text{sp}}^{\text{fin}} = 2 \times (-116\ k_B T) = -232\ k_B T$$

Thus, reproduction reduces the elastic energy by $3.68\pi\kappa$ or $116\ k_B T = 0.026\ \times\ 116$ eV = 3.02 eV. Although the bending energy increases ~10 $k_B T$ the Gaussian curvature contribution decreases with ~126 $k_B T$.

To compare this bending energy difference with the standard Gibbs free energy difference between $D^-$ in the unimer state and in the vesicle membrane state, we estimate the bending energy per molecule. The surface area of a spherical vesicle of radius 0.5 μm was calculated above to be ~ 3.14 x $10^{-12}$ $m^2$ and contains ~ 2.9 x $10^7$ molecules (0.22 $nm^2$ per decanoate). Therefore, the contribution of bending energy per molecule is on the order of $( -(116/2.9) \times 10^{-7} = -4.0 \times 10^{-6})\ k_B T$ or $-1.04 \times 10^{-7}$ eV per molecule, which is negligible compared with the free energy difference $\Delta G_f^0[D^-{}_{\text{uni}}] - \Delta G_f^0[D^-{}_{\text{ves}}] = 4.5 k_B T = 0.12$ eV per molecule.

It should be noted that for division to occur, the vesicle must not only deform into the limiting shape but also undergo neck breaking. We assume that neck breaking is achieved by the constriction force arising from spontaneous curvature [Steinkühler et al. (2020)]. The constriction force is given by:

$$f = 8\pi\kappa(C_0 - C_{ne})$$

where $C_{ne} = \frac{1}{2}\left(\frac{1}{R_1} + \frac{1}{R_2}\right)$ is the neck curvature, with $R_1$ and $R_2$ denoting the radii of the two spherical segments of the limiting-shape vesicle and $\kappa$ is the bending modulus. For symmetric division, $R_1 = R_2 =$ $R_0/\sqrt{2}$. Taking $R_0 = 0.5$ μm, $\kappa = 10\ k_BT$ and $C_0R_0 = 2.4$, the constriction force is $f \sim 2$ pN. This value is about one-tenth of the 26 pN required for breaking phospholipid membranes [Steinkühler et al. (2020)]. Since fatty acid membranes are considered easier to rupture than phospholipid membranes [Zhu & Szostak (2009)], the neck breaking is likely to occur even under this level of constriction force.

## 4.4 Self-assembly of energy-transducer and co-factor components at vesicle container

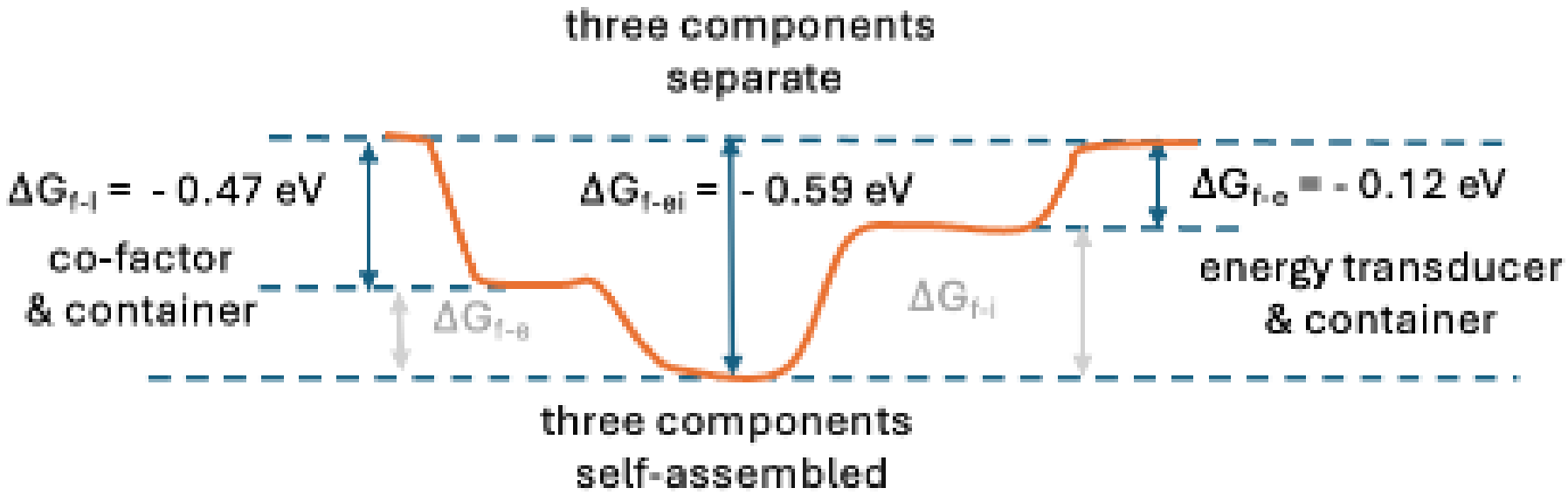

**Figure 4.4** *Free energy changes associated with co-factor (DNA) and energy transducer (RuC) component self-assembly.*

Both the RuC energy-transducer and the sdDNA co-factor are through self-assembly anchored into the vesicle. A dsDNA is about 2 nm in diameter so we may assume an anchored dsDNA occupies about 2 nm x 2 nm = 4 $nm^2$ = 4 x $10^{-18}$ $m^2$ on the vesicle surface. When also a ruthenium complex is intercalated, we can assume a bit larger area. To avoid too much crowding through dsDNA-RuC to dsDNA-RuC interactions, we may assume each dsDNA-RuC needs at least 8 nm x 8 nm = 16 $nm^2$ of free available area = 16 x $10^{-18}$ $m^2$. Such an area contains about 2 x (16 x $10^{-18}$ $m^2$ / 0.22 x $10^{-18}$ $m^2$ ) [Danov et al. (2006)] = 73 x 2 = 146 fatty acid molecules in the bilayer.

To estimate the component self-assembly energies, we need to know the details of their hydrophobic anchors. Maurer et al., (2011) provides details about a ruthenium (C-10) anchor and Wamberg et al (2014) provides details of an ssDNA (C-20 bola) anchor. This means that the maximum number of anchored dsDNA (with intercalated ruthenium complexes) on the vesicle surface is: area of vesicle / area pr dsDNA = (3.14 x $10^{-12}$ $m^2$) / (16 x $10^{-18}$ $m^2$) = 1.96 x $10^5$ ~ 2 x $10^5$.

Thus, one lifecycle needs ~ 3 x $10^7$ new lipids and ~ 2 x $10^5$ new dsDNAs, as well as 2 x $10^5$ new ruthenium complexes provided from the environment. This gives us about a 150 factor between the number of lipid molecules and the number of dsDNA together with the same number of ruthenium complexes, which we assume are intercalated with the dsDNAs at the surface.

Now, comparing the above relative number estimates with the actual metabolic reaction concentrations in previous experiments we note that the RuC and oxoG concentrations versus the fatty acid concentrations were 1 mM : 15 mM in DeClue et al. (2009) and 1 mM : 150 mM in Maurer et al. (2011). This indicates that our above estimate of 1:150 is somewhat in correspondence with the relative concentrations used in the previously implemented experimental systems.

We may estimate the free energies of the component self-assembly by using the results from our previous Subsection 4.3, where we calculated the self-assembly energies from the fatty acids were including the electrostatic potential contributions from the headgroups. Since both the ruthenium complex and the anchored dsDNA are also charged, we here approximate the free energy contributions from their anchors by counting the number of hydrophobic hydrocarbon groups.

Recall that a decanoate molecule with 9 hydrocarbons contribute with about 0.12 eV, so we may attribute 0.12 eV/9 = 0.013 eV to each hydrocarbon group. Continuing using this approximation we get the following.

Container assembly of the energy-transducer (ruthenium complex) assuming a hydrocarbon chain of length 9, the same as a decanoate, we get, $\Delta G_{f\text{-}e}$ = - 0.12 eV/molecule. Container assembly of co-factor (DNA duplex) assuming a hydrocarbon chain of 18 for each ssDNA anchor: $\Delta G_{f\text{-}i}$ = - 2 x 18 x 0.013 eV = - 0.47 eV/duplex molecule.

Thus, the self-assembly free energy from the intercalated three molecular complex (dsDNA and Ru-C): $\Delta G_{f\text{-}ei}$ = - (0.12 eV/Ru-C + 0.47 eV/dsDNA) = - 0.59 eV/component pair, and for all 2 x $10^5$ component pairs we have $\Delta G_{f\text{-}ei\text{-}m\text{-}sum}$ = 2 x $10^5$ x - 0.59 eV/component pair = - 1.18 x $10^5$ eV for all component pairs.

From this we find that the free energy of self-assembly for the new vesicle is about 30 times larger than the self-assembly energy of the new co-factor and energy-transducer component pairs.

## 4.5 Energetics of a protocellular lifecycle

We can now summarize the free energy estimates associated with a full lifecycle. This is the externally provided free energies associated with the production of the needed building blocks; the important self-assembly energies that ensures the functional components are kept together and thus enables the protocell to obtain its key properties including self-maintenances, growth and replication, where replication is enabled by the weak membrane bending energies.

We may treat separately the free energies of formation $\Delta G_{metbolism}$, and the free energies stored in the system from self-assembly $\Delta G_{selfassembly}$ as the $\Delta G_{metbolism}$ is generated by mainly irreversible nonequilibrium processes, while the self-assembly $\Delta G_{selfassembly}$ (and the bending $\Delta G_{bening}$) energies are determined by reversible equilibrium (downhill) processes.

| | |
|---|---|
| Free energy input for formation of a new fatty acid molecule | + 2.1 eV/fatty acid |
| Free energy change from input to output for fatty acid formation | - 1.62 eV/fatty acid |
| Free energy input for formation of a new dsDNA co-factor | + 4.88 eV/dsDNA |
| Free energy change from metabolic co-factor input to output | - 0.98 eV/dsDNA |
| dsDNA hybridization energy per new dsDNA | - 0.56 eV/dsDNA |
| Ruthenium complex intercalation within dsDNA | - 0.24 eV/intercalation |
| Container assembly of vesicle/lipid per molecule | - 0.12 eV/molecule |
| Container assembly of energy transducer and co-factor catalyst pair | - 0.59 eV/catalyst pair |
| Container bending energy per lipid molecule in vesicle (vesicle fission) | - $1.04 \times 10^{-7}$ eV/molecule |

***Table 4.5.*** *Summary of the free energy hierarchy for the protocellular lifecycle.*

The free energy changes for the metabolism in the protocellular lifecycle stems from two interconnected processes. One accounts for the direct free energy input in terms of photo energy and the other from input in terms of energized bonds that accounts for the free energy changes from reactants (resources) to products (building blocks) in the metabolic processes. Thus, this protocell both have autotropic and heterotopic features.

Free energy input from photo energy and ligation activation (chemical bond energy) for each lifecycle:

$$\Delta G_{metabolism_1} = 2 \text{ x } 10^5 \text{ x } [150 \text{ lipids} + 1 \text{ dsDNA}]_{input} =$$
$$= 2 \text{ x } 10^5 \text{ x } (150 \text{ x } (+2.1 \text{ eV}) + 4.88 \text{ eV})$$
$$= + 6.4 \text{ x}10^7 \text{ eV or } 1.02 \text{ x}10^{-11} \text{ J.}$$

Free energy changes from internal chemical bond energy in resources that is released in the production of building blocks for each lifecycle:

$$\Delta G_{metabolism_2} = 2 \text{ x } 10^5 \text{ x } [150 \text{ lipids} + 1 \text{ dsDNA}]_{resources} =$$
$$= 2 \text{ x } 10^5 \text{ x } (150 \text{ x } (-1.62 \text{ eV}) - 0.98 \text{ eV})$$
$$= - 4.88 \text{ x}10^7 \text{ eV or } - 7.82 \text{ x}10^{-12} \text{ J.}$$

The free energy is mainly used to produce fatty acids. The co-factor DNA production requires less than 1% of the available energy.

For a lifecycle we can now estimate the total metabolic input of free energy both from light and from chemical bond energy within the resources:

$$\Delta G_{metbolism} = \Delta G_{metabolism_1} - \Delta G_{metabolism_2} = 6.4 \text{ x}10^7 \text{ eV} + 4.88 \text{ x}10^7 \text{ eV} = 1.13 \text{ x}10^8 \text{ eV or } 1.81 \text{ x}10^{-11} \text{ J.}$$

The free energy changes due to the self-assembly processes can be estimated as:

$$\Delta G_{selfassembly} = 1 \text{ vesicle} + 2 \text{ x } 10^5 \text{ x (energy-transducer and co-factor anchors)}$$
$$= - 3.48 \text{ x } 10^6 \text{ eV/vesicle} + 2 \text{ x } 10^5 \text{ (- 0.59 eV/catalyst pair)}$$
$$= - 3.72 \text{ x } 10^6 \text{ eV or } - 5.96 \text{ x}10^{-13} \text{ J.}$$

Note that $dG_{metbolism}$ is positive (added to the system) and numerically about ~ 30 times larger than $\Delta G_{selfassembly}$ that is negative and stored within the system. In Section 4.3 we estimated the total free energy gain form the vesicle fission to be about 3.02 eV, so we can ignore the bending energies in these estimations as they are about 6 orders of magnitude smaller than the self-assembly energies and more than 7 orders of magnitude smaller than the metabolic energy transformations.

Thus, a numerical value estimate of all free energy *changes* in a protocellular lifecycle, input (positive) and internal (negative) energies are:

$$|\Delta G_{metabolism}| + |\Delta G_{selfassembly}| = 1.13 \text{ x}10^8 \text{ eV} + 3.72 \text{ x}10^6 \text{ eV} = 1.17 \text{ x}10^8 \text{ eV or } 1.87 \text{ x}10^{-11} \text{ J.}$$

This is summarized graphically in Fig. 4.5.

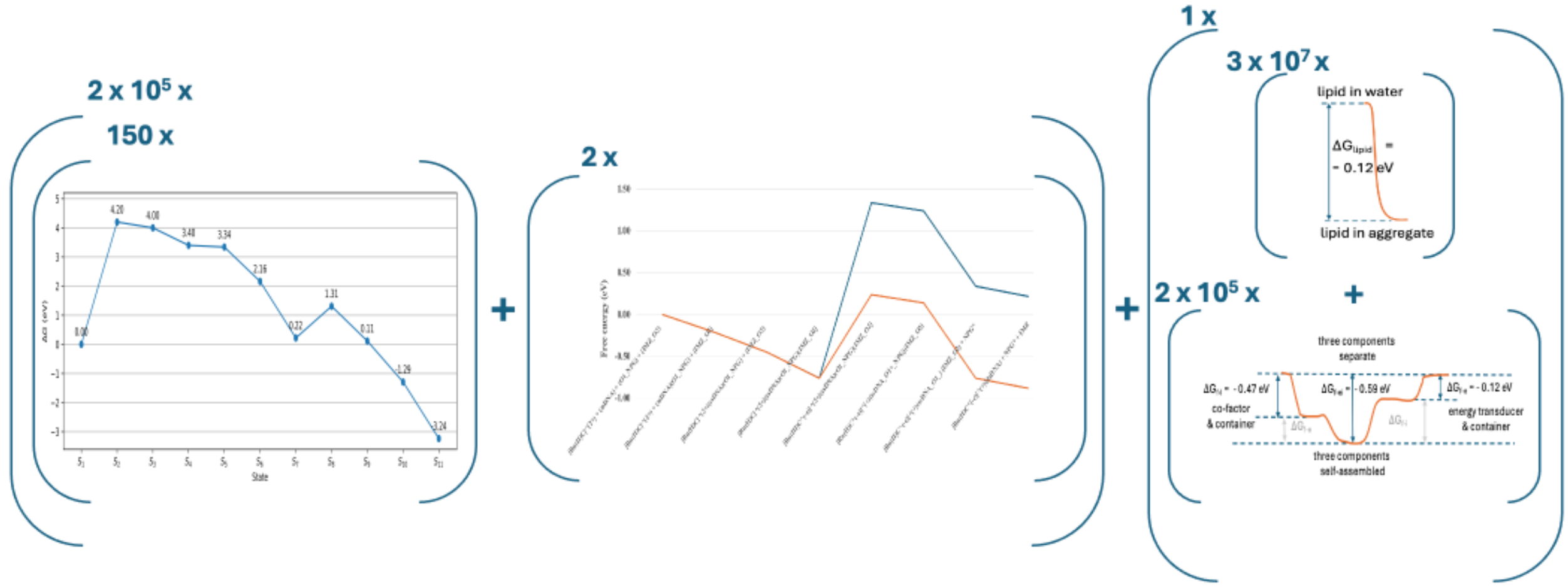

**Figure 4.5** *Energetic summary depiction of the protocellular lifecycle. The numbers at the upper left corner of the parentheses refer to the approximate number of the associated reactions involved in one complete lifecycle. Starting with the large parenthesis to the left that accounts for the metabolic production: The left smaller parenthesis inside the large left parenthesis indicates that 2 x $10^5$ x 150 new fatty acids are produced through photo activation and digestion of resource molecules. See subsection 4.1 for details. The next small parenthesis to the right inside the large left parenthesis indicates that 2 x $10^5$ x 2 new DNA co-factor strands are produced, where the DNA replication involves photo activation and DNA charge transfer combined with a chemical activation that results in oligomer ligation. This parenthesis also includes a depiction of the intermediate reversable hybridization energies that are not included in this estimation. See subsection 4.2 for details. For simplicity we assume only one (1 x) vesicle division occurs in each lifecycle (large parenthesis to the right), which involves the self-assembly of about 2 x $10^7$ newly formed lipids, see subsection 4.3 for details, as well as anchoring of about 2 x $10^5$ ruthenium complex energy transducers and DNA co-factors, see subsection 4.4 for details. This parenthesis accounts for the changes in the protocellular self-assembly energies. Note that for this summary we are ignore the vesicle bending energies as they are about $10^6$ times smaller than the self-assembly energies.*

# 5 Discussion

What are the main difference and similarities between modern unicellular metabolisms and the metabolism of the analyzed protocell? There are multiple significant differences between the protocellular metabolism and the metabolism of a modern cell, where the most striking difference in complexity between the two.

Another significant difference is the lack of a universal energy storage and transporter molecule as modern life ATP as well as proton and sodium gradients across a cell membrane. Despite these differences, simple light-driven electron transport systems can drive the protocelular metabolism.
The quantum free energy efficiency of the protocell is ~ 0.5% (of the initial photon energy that enters the metabolic pathways to produce building blocks, [DeClue et al. (2009]) while modern photosynthesis has an efficiency of ~ 30% (photon energy to glucose), with an overall energy efficiency of a green plant between 0.1 and 1%.

Yet another difference, the analyzed protocell has no translation machinery and use DNA as an integral part of the metabolism through electron donation and charge transport. It is noteworthy that the modern energy transporter ATP is structurally very similar to the DNA nucleobases, which might be an echo from an early evolutionary connection between energy transducer and information [Shelly et al., (2007)].

We believe, however, that there are some deeper and perhaps not immediately obvious similarities between contemporary unicells and the protocell. Dihydrophenylglycine acts as a proton donor in a similar manner as water does in modern photosynthesis. Both metabolisms have electron transport chains, and both use light as an energy source. We may also compare the terminal electron acceptor in picolinium ester of a fatty acid in the protocell as opposed to NADP+ in a modern cell. Expressed differently, our pathway produces fat (fatty acid) instead of sugar (glucose). The ruthenium complex captures photons, analogous to chlorophyll, so one may we view $(bpy)_3$ as corresponding to the porphyrin ring that holds the $Mg^{2+}$ of chlorophyll. In this case $Ru^{2+}$ would be in place of $Mg^{2+}$. These comparisons are not perfect, but we believe they can help provide a bridge for a better understanding of both systems.

Finally, we may compare the protocellular metabolism with the observed biological scaling laws for modern unicells [West & Brown (2005)]. We can calculate the metabolic energy per produced wet biomass weight (including water). Assuming a one-micron protocell we have the metabolic energy of $1.41 \times 10^{-11}$ J and a volume = $4/3\ \pi\ (0.5 \times 10^{-6}\ \text{m})^3 = 5.24 \times 10^{-13}\ \text{cm}^3$. As the protocellular volume is dominated by the water in the vesicle lumen, we may approximate the weight of whole protocell with the weight of a droplet of water with the same diameter (and a density of 1 $\text{g/cm}^3$), which yields 27 J/g. This, however, is not a good indicator for the protocellular energy efficiency, as only the molecular complexes at the vesicle surface are involved in the metabolic processes while the vesicle lumen is passive.

The energy per produced dry weight can be calculated from the molecular numbers provided in Section 4.5. Here we get $2 \times 10^5 \times 150$ decanoic acid molecules that weigh $8.6 \times 10^{-15}$ g and $2 \times 10^5$ dsDNA with 18 base pairs each with C-20 hydrophobic anchors that weigh about $4 \times 10^{-14}$ g, which gives a combined dry weight of about $4.9 \times 10^{-14}$ g. Note that the organometallic ruthenium complexes are provided by the environment so they should not be included in this estimation. Thus, the metabolic energy per produced dry weight of biomass is 288 J/g.

For comparison, Ortega-Arzola et al. (2024) report an energy efficiency of a bit more than 300 J/g dry weight for unicellular organisms, which is very similar to that of the investigated protocell. As for modern cell, the protocell uses most of its free energy (~ 99%) to generate membrane molecules.

# 6. Conclusion

We have presented a method to calculate the hierarchy of all free energy changes in a lifecycle in a simple integrated protocellular system where metabolism, information and container are autocatalytically coupled. The presented approach can also be used to estimate free energy changes in other protocellular or simple modern biological systems.

The free energy changes in this system range from the metabolic reactions that involves the formation and breakage of chemical bonds including information replication, over component and vesicle self-assembly processes, as well as the weak energies responsible for membrane bending and vesicle self-reproduction. The metabolic reactions accounts for about 97% of all free energy changes in the lifecycle, while the self-assembly free energy changes account for about 3%. The membrane bending energies are five orders of magnitude smaller than the metabolic energies so we can neglect these energies in this analysis. Nerveless, these tiny energies are crucial for vesicle reproduction and thus for the protocellular replication.

The investigated protocell is both fueled by photo energy and chemical bond energy, so its metabolism is neither like a modern autotroph nor like a modern heterotroph; it is a mixture of both. If we view this in an origin of life perspective, the first life forms may not either have been autolithotrophs, as the young Earth presumably had a richer composition of complex organics than what we find today. If that is true, it seems reasonable to assume that the first life forms would have utilized the simplest possible metabolic processes and not at the onset developed more sophisticated metabolisms. Later life may have depleted this initial easy food, which presumably would have forced more efficient metabolisms to evolve in response to this environmental change.

At an aggregate level, it should be noted that the presented protocellular photo-driven part of the metabolism does not support metabolic reactions for end products with higher free energy than the input resource reactants. That is not the case e.g. for modern photosynthesis where water and carbon dioxide are transformed into sucker (organics). To our knowledge, this is an issue for all published integrated protocelluar metabolisms. They miss the ability to support covalent bond formation to create complex building blocks from simple resource molecules.

For a protocell of diameter 1 micron the estimated free energy changes of all processes are about 1.17 $x10^{8}$ eV or 1.87 $x10^{-11}$ J, while only the metabolic changes account for about 1.13 $x10^{8}$ eV or 1.81 $x10^{-11}$ J. In comparison, this is about 14 times smaller than the free energy change in the lifecycle for a E. coli of similar size, which is estimated to be about $2.0 \times 10^{-10}$ J or $1.2 \times 10^{9}$ eV. We should expect that for similar sizes, the metabolic transformation energies are significantly lower for the investigated protocell. Its metabolism is limited to the (2D) vesicle surface, while the E. coli has its metabolism within its lumen (3D). Further, we find a protocellular metabolic efficiency of about 300 J/g dry biomass, which is similar to the reported efficiency of modern unicells. Also, as modern unicells, the investigated protocell uses most of its free energy to produce membrane container molecules.

# APPENDIX / Supplementary Information

## SI.1. DFT calculations of gas-phase free energies involved in the metabolic processes that produce fatty acid

Gas-phase DFT calculations were performed for several molecular species involved in the overall reactions using the ORCA code [Neese et al. (2020)]. The B3LYP+D3BJ functional in combination with the def2-TZVP basis set was employed as our reference method. Gibbs free energies (G) at 298 K were obtained from ORCA frequency calculations as the sum of the electronic energy and the thermal free-energy correction. This means that the obtained free energies are calculated from a hypothetical state where all electrons and nucleus are infinitely far apart (U = G = 0) to the equilibrium configuration of each of the molecules. A negative energy means that the given molecule is in a stable configuration. Although solvent effects are not included, the computed gas-phase energies provide qualitative insight into the intrinsic thermodynamic driving forces that are expected to persist in solution. The calculated internal and free energies are listed in table SI1.1 below.

| | **B3LYP+D3BJ** | | **B3LYP** | | **PBE** | | **PBE+D3BJ** | | **wB97x** | |
|---|---|---|---|---|---|---|---|---|---|---|
| **Fragment** | *U (eV)* | *G (eV)* | *U (eV)* | *G (eV)* | *U (eV)* | *G (eV)* | *U (eV)* | *G (eV)* | *U (eV)* | *G (eV)* |
| oxoG | **-16809.535** | **-16808.078** | -16809.535 | -16807.202 | -16799.963 | -16797.726 | -16799.963 | -16798.241 | -16814.028 | -16812.571 |
| oxoG+* | **-16802.331** | **-16800.851** | -16802.331 | -16799.972 | -16792.810 | -16790.550 | -16792.810 | -16791.067 | -16806.694 | -16805.214 |
| oxoG* | **-16792.043** | **-16790.908** | -16792.044 | -16790.070 | -16782.552 | -16780.669 | -16782.552 | -16781.160 | -16796.364 | -16795.229 |
| HDH2 | **-14056.295** | **-14053.368** | -14056.298 | -14052.291 | -14047.374 | -14043.520 | -14047.373 | -14044.161 | -14060.878 | -14057.951 |
| HDH* | **-14039.366** | **-14036.787** | -14039.369 | -14035.740 | -14030.520 | -14027.027 | -14030.519 | -14027.650 | -14043.748 | -14041.169 |
| HD | **-14024.562** | **-14022.148** | -14024.564 | -14021.134 | -14015.728 | -14012.427 | -14015.727 | -14013.026 | -14028.799 | -14026.385 |
| Lp+ | **-24728.342** | **-24719.613** | -24728.345 | -24717.391 | -24711.504 | -24700.902 | -24711.502 | -24702.248 | -24737.568 | -24728.839 |
| Lp* | **-24733.206** | **-24724.595** | -24733.209 | -24722.386 | -24716.412 | -24705.940 | -24716.411 | -24707.275 | -24742.494 | -24733.883 |
| L | **-15857.527** | **-15851.443** | -15857.529 | -15850.100 | -15846.595 | -15839.404 | -15846.594 | -15840.223 | -15863.627 | -15857.543 |
| p+* | **-8886.019** | **-8883.723** | -8886.019 | -8882.964 | -8879.904 | -8876.954 | -8879.904 | -8877.401 | -8888.934 | -8886.638 |
| pH+ | **-8903.670** | **-8901.007** | -8903.671 | -8900.226 | -8897.524 | -8894.196 | -8897.523 | -8894.657 | -8906.749 | -8904.086 |

***Table SI1.1*** *Result of DFT internal energy U and free energy G at 298 K computed with ORCA for the molecular moieties in Fig. 4.1.3 using different exchange-correlation functionals and/or dispersion correction.*

We also compared our results using different exchange-correlation functionals and/or dispersion correction, and found that our reaction energies are qualitatively robust, see Fig. SI1.1.

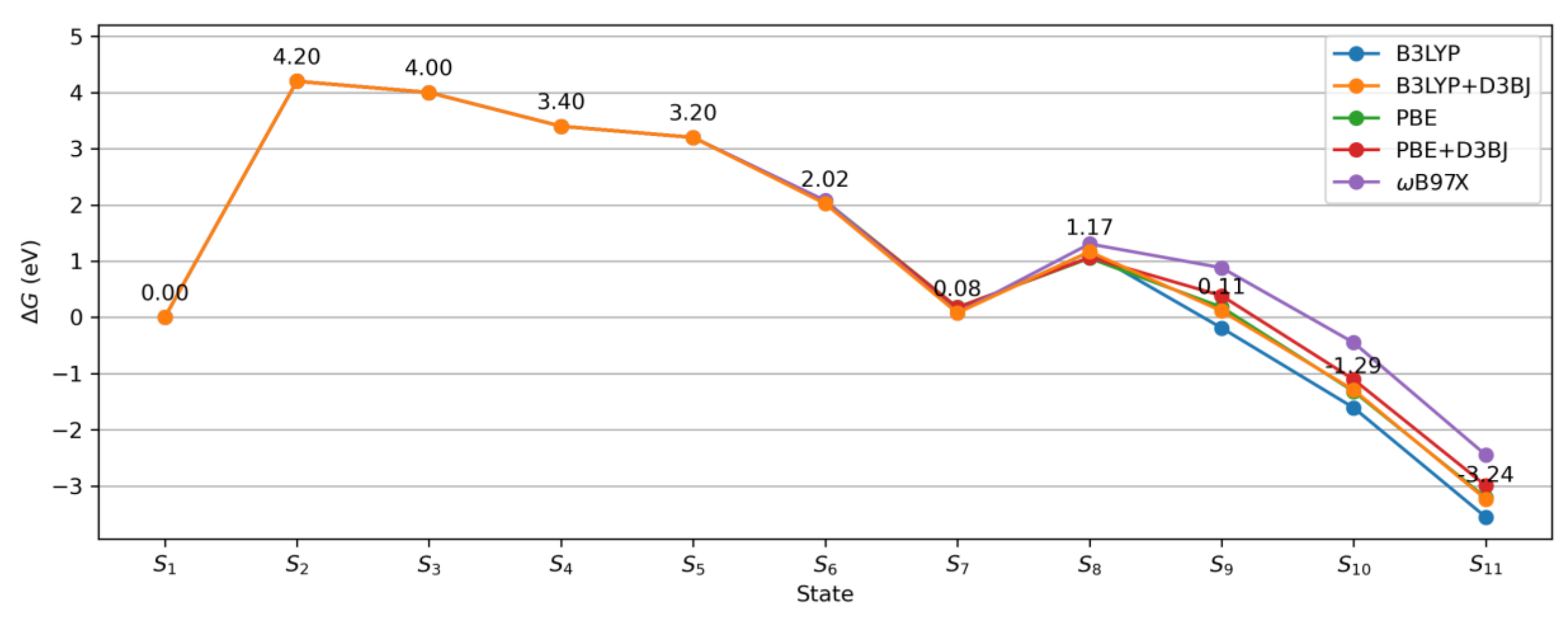

***Fig. SI1.1*** *Comparison of the computed overall reaction pathway based on data from different approximations for the DFT input, cf. Tab. SI1.1. Note that the results are qualitative robust with respect to different exchange-correlation functionals and/or dispersion corrections.*

## SI.2 Equation system for calculating remaining unknown free energy changes involved in the metabolic processes that produce fatty acid

We have determined 11 states that the system can be in (S1, … S11) and we here present a set of linear equations that enable to solve the corresponding free energies with various information at hand. The coefficient matrix for this linear equation system is shown in Fig SI2.1

B3LYP-D3BJ

| Ru(II)C2+ | Ru(III)Ce2+ | Ru(II)Ce+ | dsDNA | dsDNA+ | oxoG | oxoG+* | oxoG* | HDH2 | HDH* | HD | Lp+ | Lp* | L | p+* | pH+ | H+ | e | S1 | S2 | S3 | S4 | S5 | S6 | S7 | S8 | S9 | S10 | S11 | Description of information Encoded in the matrix row | Constants (eV) | Result (eV) |
|---|---|---|---|---|---|---|---|---|---|---|---|---|---|---|---|---|---|---|---|---|---|---|---|---|---|---|---|---|---|---|---|
| 2 | 0 | 0 | 2 | 0 | 2 | 0 | 0 | 3 | 0 | 0 | 2 | 0 | 0 | 0 | 0 | 0 | 0 | -1 | 0 | 0 | 0 | 0 | 0 | 0 | 0 | 0 | 0 | 0 | detailed balance | 0.00 | -50000.00 |
| 0 | 2 | 0 | 2 | 0 | 2 | 0 | 0 | 3 | 0 | 0 | 2 | 0 | 0 | 0 | 0 | 0 | 0 | 0 | -1 | 0 | 0 | 0 | 0 | 0 | 0 | 0 | 0 | 0 | detailed balance | 0.00 | -49997.90 |
| 0 | 0 | 2 | 0 | 2 | 2 | 0 | 0 | 3 | 0 | 0 | 2 | 0 | 0 | 0 | 0 | 0 | 0 | 0 | 0 | -1 | 0 | 0 | 0 | 0 | 0 | 0 | 0 | 0 | detailed balance | 0.00 | -50005.53 |
| 0 | 0 | 2 | 2 | 0 | 0 | 2 | 0 | 3 | 0 | 0 | 2 | 0 | 0 | 0 | 0 | 0 | 0 | 0 | 0 | 0 | -1 | 0 | 0 | 0 | 0 | 0 | 0 | 0 | detailed balance | 0.00 | -60000.00 |
| 0 | 0 | 2 | 2 | 0 | 0 | 0 | 2 | 3 | 0 | 0 | 2 | 0 | 0 | 0 | 0 | 2 | 0 | 0 | 0 | 0 | 0 | -1 | 0 | 0 | 0 | 0 | 0 | 0 | detailed balance | 0.00 | -59992.47 |
| 0 | 0 | 2 | 2 | 0 | 2 | 0 | 0 | 1 | 2 | 0 | 2 | 0 | 0 | 0 | 0 | 2 | 0 | 0 | 0 | 0 | 0 | 0 | -1 | 0 | 0 | 0 | 0 | 0 | detailed balance | 0.00 | -16808.08 |
| 0 | 0 | 2 | 2 | 0 | 2 | 0 | 0 | 2 | 0 | 1 | 2 | 0 | 0 | 0 | 0 | 2 | 0 | 0 | 0 | 0 | 0 | 0 | 0 | -1 | 0 | 0 | 0 | 0 | detailed balance | 0.00 | -16800.85 |
| 2 | 0 | 0 | 2 | 0 | 2 | 0 | 0 | 2 | 0 | 1 | 0 | 2 | 0 | 0 | 0 | 2 | 0 | 0 | 0 | 0 | 0 | 0 | 0 | 0 | -1 | 0 | 0 | 0 | detailed balance | 0.00 | -16790.91 |
| 2 | 0 | 0 | 2 | 0 | 2 | 0 | 0 | 2 | 0 | 1 | 0 | 0 | 2 | 2 | 0 | 0 | 0 | 0 | 0 | 0 | 0 | 0 | 0 | 0 | 0 | -1 | 0 | 0 | detailed balance | 0.00 | -14053.37 |
| 2 | 0 | 0 | 2 | 0 | 2 | 0 | 0 | 0 | 2 | 1 | 0 | 0 | 2 | 0 | 2 | 0 | 0 | 0 | 0 | 0 | 0 | 0 | 0 | 0 | 0 | 0 | -1 | 0 | detailed balance | 0.00 | -14036.79 |
| 2 | 0 | 0 | 2 | 0 | 2 | 0 | 0 | 1 | 0 | 2 | 0 | 0 | 2 | 0 | 2 | 0 | 0 | 0 | 0 | 0 | 0 | 0 | 0 | 0 | 0 | 0 | 0 | -1 | detailed balance | 0.00 | -14022.15 |
| 0 | 0 | 0 | 0 | 0 | 1 | 0 | 0 | 0 | 0 | 0 | 0 | 0 | 0 | 0 | 0 | 0 | 0 | 0 | 0 | 0 | 0 | 0 | 0 | 0 | 0 | 0 | 0 | 0 | ORCA DFT | -16808.08 | -24719.61 |
| 0 | 0 | 0 | 0 | 0 | 0 | 1 | 0 | 0 | 0 | 0 | 0 | 0 | 0 | 0 | 0 | 0 | 0 | 0 | 0 | 0 | 0 | 0 | 0 | 0 | 0 | 0 | 0 | 0 | ORCA DFT | -16800.85 | -24724.59 |
| 0 | 0 | 0 | 0 | 0 | 0 | 0 | 1 | 0 | 0 | 0 | 0 | 0 | 0 | 0 | 0 | 0 | 0 | 0 | 0 | 0 | 0 | 0 | 0 | 0 | 0 | 0 | 0 | 0 | ORCA DFT | -16790.91 | -15851.44 |
| 0 | 0 | 0 | 0 | 0 | 0 | 0 | 0 | 1 | 0 | 0 | 0 | 0 | 0 | 0 | 0 | 0 | 0 | 0 | 0 | 0 | 0 | 0 | 0 | 0 | 0 | 0 | 0 | 0 | ORCA DFT | -14053.37 | -8883.72 |
| 0 | 0 | 0 | 0 | 0 | 0 | 0 | 0 | 0 | 1 | 0 | 0 | 0 | 0 | 0 | 0 | 0 | 0 | 0 | 0 | 0 | 0 | 0 | 0 | 0 | 0 | 0 | 0 | 0 | ORCA DFT | -14036.79 | -8901.01 |
| 0 | 0 | 0 | 0 | 0 | 0 | 0 | 0 | 0 | 0 | 1 | 0 | 0 | 0 | 0 | 0 | 0 | 0 | 0 | 0 | 0 | 0 | 0 | 0 | 0 | 0 | 0 | 0 | 0 | ORCA DFT | -14022.15 | -9.97 |
| 0 | 0 | 0 | 0 | 0 | 0 | 0 | 0 | 0 | 0 | 0 | 1 | 0 | 0 | 0 | 0 | 0 | 0 | 0 | 0 | 0 | 0 | 0 | 0 | 0 | 0 | 0 | 0 | 0 | ORCA DFT | -24719.61 | -6.63 |
| 0 | 0 | 0 | 0 | 0 | 0 | 0 | 0 | 0 | 0 | 0 | 0 | 1 | 0 | 0 | 0 | 0 | 0 | 0 | 0 | 0 | 0 | 0 | 0 | 0 | 0 | 0 | 0 | 0 | ORCA DFT | -24724.59 | -345215.49 |
| 0 | 0 | 0 | 0 | 0 | 0 | 0 | 0 | 0 | 0 | 0 | 0 | 0 | 1 | 0 | 0 | 0 | 0 | 0 | 0 | 0 | 0 | 0 | 0 | 0 | 0 | 0 | 0 | 0 | ORCA DFT | -15851.44 | -345211.29 |
| 0 | 0 | 0 | 0 | 0 | 0 | 0 | 0 | 0 | 0 | 0 | 0 | 0 | 0 | 1 | 0 | 0 | 0 | 0 | 0 | 0 | 0 | 0 | 0 | 0 | 0 | 0 | 0 | 0 | ORCA DFT | -8883.72 | -345211.49 |
| 0 | 0 | 0 | 0 | 0 | 0 | 0 | 0 | 0 | 0 | 0 | 0 | 0 | 0 | 0 | 1 | 0 | 0 | 0 | 0 | 0 | 0 | 0 | 0 | 0 | 0 | 0 | 0 | 0 | ORCA DFT | -8901.01 | -345212.09 |
| -1 | 1 | 0 | 0 | 0 | 0 | 0 | 0 | 0 | 0 | 0 | 0 | 0 | 0 | 0 | 0 | 0 | 0 | 0 | 0 | 0 | 0 | 0 | 0 | 0 | 0 | 0 | 0 | 0 | light excitation | **2.10** | -345212.15 |
| 0 | -1 | 1 | 0 | 0 | 0 | 0 | 0 | 0 | 0 | 0 | 0 | 0 | 0 | 0 | 0 | 0 | -1 | 0 | 0 | 0 | 0 | 0 | 0 | 0 | 0 | 0 | 0 | 0 | Ru electron capture | **-1.00** | -345213.33 |
| 0 | 0 | 0 | 1 | -1 | -1 | 1 | 0 | 0 | 0 | 0 | 0 | 0 | 0 | 0 | 0 | 0 | 0 | 0 | 0 | 0 | 0 | 0 | 0 | 0 | 0 | 0 | 0 | 0 | hole diffusion | **-0.30** | -345215.27 |
| 0 | 0 | 0 | 0 | 0 | -1 | 1 | 0 | 0 | 0 | 0 | 0 | 0 | 0 | 0 | 0 | 0 | 1 | 0 | 0 | 0 | 0 | 0 | 0 | 0 | 0 | 0 | 0 | 0 | oxidation of oxoG (e-removal) | **0.60** | -345214.18 |
| 0 | 0 | 0 | 0 | 0 | 0 | -1 | 1 | 0 | 0 | 0 | 0 | 0 | 0 | 0 | 0 | 1 | 0 | 0 | 0 | 0 | 0 | 0 | 0 | 0 | 0 | 0 | 0 | 0 | Spontaneous | **-0.03** | -345215.37 |
| 1 | 0 | 0 | 0 | 0 | 0 | 0 | 0 | 0 | 0 | 0 | 0 | 0 | 0 | 0 | 0 | 0 | 0 | 0 | 0 | 0 | 0 | 0 | 0 | 0 | 0 | 0 | 0 | 0 | nom assignment | **-50000.00** | -345216.78 |
| 0 | 0 | 0 | 1 | 0 | 0 | 0 | 0 | 0 | 0 | 0 | 0 | 0 | 0 | 0 | 0 | 0 | 0 | 0 | 0 | 0 | 0 | 0 | 0 | 0 | 0 | 0 | 0 | 0 | nom assignment | **-60000.00** | -345218.72 |

**Fig. SI.2.1** *Coefficient matrix* ***A*** *(left box) for 29 equations with 29 free energy parameters* ***x*** *(18 molecular species and 11 overall reaction states $S_1$, … $S_{11}$). In the right box the constants (**y**) for the equations are listed as well as the resulting solution $\mathbf{x} = \mathbf{A}^{-1}\,\mathbf{y}$, thus providing the free energies of all the involved species (orange block) and the free energies of the states $S_1$, … $S_{11}$ (red block). Regarding the constants in the blue block, we have introduced values from the literature and guesstimated the remaining two. Electron capture by the excited ruthenium complex is known to be - 1.0 eV, [Vlcek et al. (1995)], while the oxidation potential of oxoguanine is + 0.6 eV [DeClue et al. (2009)]. For the hole diffusion in the dsDNA from the formation at the original guanine site till it reaches the absorbing oxoguanine site, we guesstimate - 0.3 eV. For the process where the oxoguanine loses a hydrogen and becomes a radical (after the oxoguanine has donated an electron) we have set the free energy change to - 0.03 eV. It turns out the exact value of these two free energy changes have very little numerical impact on the resulting free energy changes. Up to 100% value changes do not change the qualitative free energy landscape for the overall processes.*

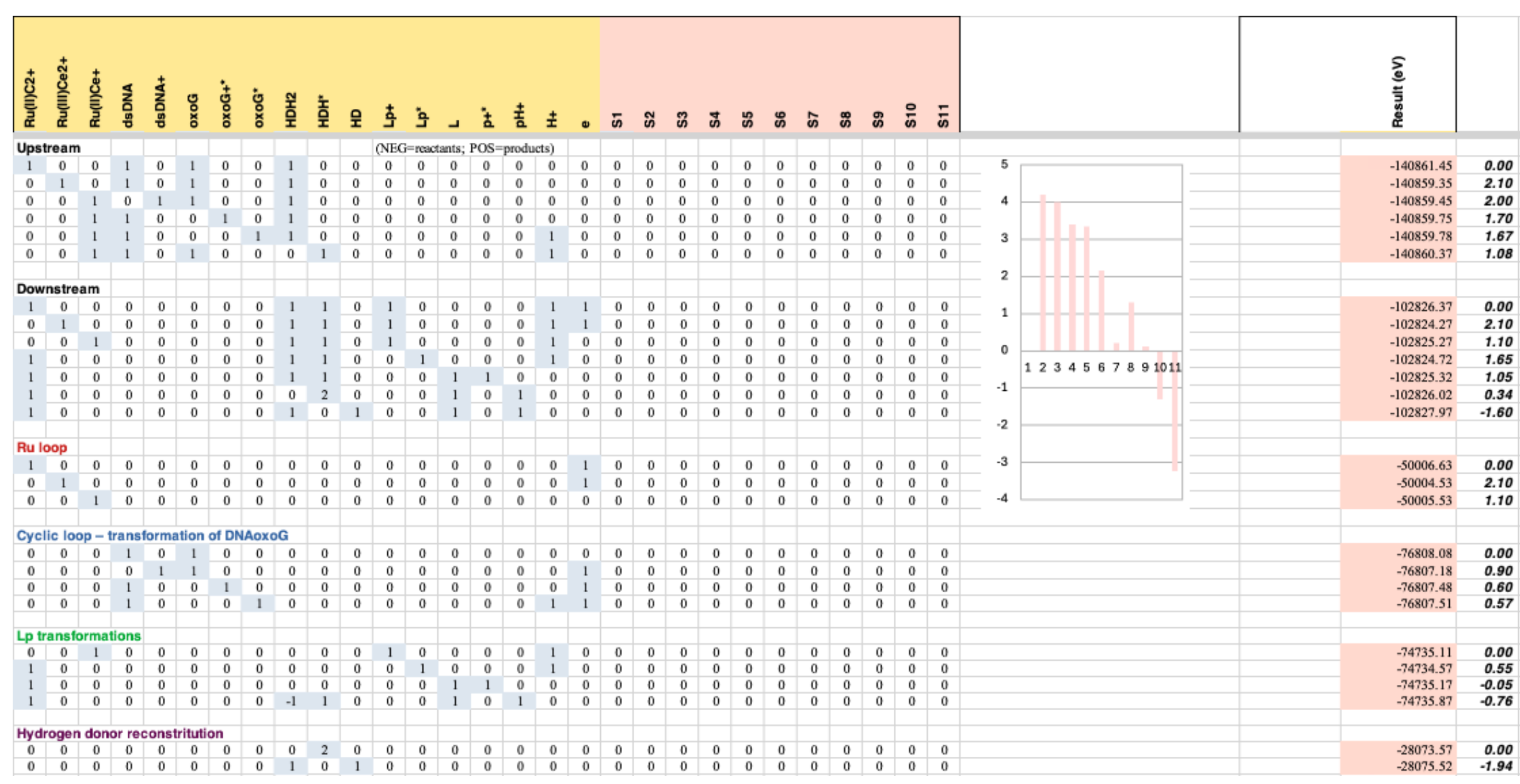

| Ru(II)C2+ | Ru(III)Ce2+ | Ru(II)Ce+ | dsDNA | dsDNA+ | oxoG | oxoG+* | oxoG* | HDH2 | HDH* | HD | Lp+ | Lp* | L | p+* | pH+ | H+ | e | S1 | S2 | S3 | S4 | S5 | S6 | S7 | S8 | S9 | S10 | S11 | Result (eV) | |
|---|---|---|---|---|---|---|---|---|---|---|---|---|---|---|---|---|---|---|---|---|---|---|---|---|---|---|---|---|---|---|
| **Upstream** | | | | | | | | | | | (NEG=reactants; POS=products) | | | | | | | | | | | | | | | | | | | |
| 1 | 0 | 0 | 1 | 0 | 1 | 0 | 0 | 1 | 0 | 0 | 0 | 0 | 0 | 0 | 0 | 0 | 0 | 0 | 0 | 0 | 0 | 0 | 0 | 0 | 0 | 0 | 0 | 0 | -140861.45 | ***0.00*** |
| 0 | 1 | 0 | 1 | 0 | 1 | 0 | 0 | 1 | 0 | 0 | 0 | 0 | 0 | 0 | 0 | 0 | 0 | 0 | 0 | 0 | 0 | 0 | 0 | 0 | 0 | 0 | 0 | 0 | -140859.35 | ***2.10*** |
| 0 | 0 | 1 | 0 | 1 | 1 | 0 | 0 | 1 | 0 | 0 | 0 | 0 | 0 | 0 | 0 | 0 | 0 | 0 | 0 | 0 | 0 | 0 | 0 | 0 | 0 | 0 | 0 | 0 | -140859.45 | ***2.00*** |
| 0 | 0 | 1 | 1 | 0 | 0 | 1 | 0 | 1 | 0 | 0 | 0 | 0 | 0 | 0 | 0 | 0 | 0 | 0 | 0 | 0 | 0 | 0 | 0 | 0 | 0 | 0 | 0 | 0 | -140859.75 | ***1.70*** |
| 0 | 0 | 1 | 1 | 0 | 0 | 0 | 1 | 1 | 0 | 0 | 0 | 0 | 0 | 0 | 0 | 1 | 0 | 0 | 0 | 0 | 0 | 0 | 0 | 0 | 0 | 0 | 0 | 0 | -140859.78 | ***1.67*** |
| 0 | 0 | 1 | 1 | 0 | 1 | 0 | 0 | 0 | 1 | 0 | 0 | 0 | 0 | 0 | 0 | 1 | 0 | 0 | 0 | 0 | 0 | 0 | 0 | 0 | 0 | 0 | 0 | 0 | -140860.37 | ***1.08*** |
| **Downstream** | | | | | | | | | | | | | | | | | | | | | | | | | | | | | | |
| 1 | 0 | 0 | 0 | 0 | 0 | 0 | 0 | 1 | 1 | 0 | 1 | 0 | 0 | 0 | 0 | 1 | 1 | 0 | 0 | 0 | 0 | 0 | 0 | 0 | 0 | 0 | 0 | 0 | -102826.37 | ***0.00*** |
| 0 | 1 | 0 | 0 | 0 | 0 | 0 | 0 | 1 | 1 | 0 | 1 | 0 | 0 | 0 | 0 | 1 | 1 | 0 | 0 | 0 | 0 | 0 | 0 | 0 | 0 | 0 | 0 | 0 | -102824.27 | ***2.10*** |
| 0 | 0 | 1 | 0 | 0 | 0 | 0 | 0 | 1 | 1 | 0 | 1 | 0 | 0 | 0 | 0 | 1 | 0 | 0 | 0 | 0 | 0 | 0 | 0 | 0 | 0 | 0 | 0 | 0 | -102825.27 | ***1.10*** |
| 1 | 0 | 0 | 0 | 0 | 0 | 0 | 0 | 1 | 1 | 0 | 0 | 1 | 0 | 0 | 0 | 1 | 0 | 0 | 0 | 0 | 0 | 0 | 0 | 0 | 0 | 0 | 0 | 0 | -102824.72 | ***1.65*** |
| 1 | 0 | 0 | 0 | 0 | 0 | 0 | 0 | 1 | 1 | 0 | 0 | 0 | 1 | 1 | 0 | 0 | 0 | 0 | 0 | 0 | 0 | 0 | 0 | 0 | 0 | 0 | 0 | 0 | -102825.32 | ***1.05*** |
| 1 | 0 | 0 | 0 | 0 | 0 | 0 | 0 | 0 | 2 | 0 | 0 | 0 | 1 | 0 | 1 | 0 | 0 | 0 | 0 | 0 | 0 | 0 | 0 | 0 | 0 | 0 | 0 | 0 | -102826.02 | ***0.34*** |
| 1 | 0 | 0 | 0 | 0 | 0 | 0 | 0 | 1 | 0 | 1 | 0 | 0 | 1 | 0 | 1 | 0 | 0 | 0 | 0 | 0 | 0 | 0 | 0 | 0 | 0 | 0 | 0 | 0 | -102827.97 | ***-1.60*** |
| **Ru loop** | | | | | | | | | | | | | | | | | | | | | | | | | | | | | | |
| 1 | 0 | 0 | 0 | 0 | 0 | 0 | 0 | 0 | 0 | 0 | 0 | 0 | 0 | 0 | 0 | 0 | 1 | 0 | 0 | 0 | 0 | 0 | 0 | 0 | 0 | 0 | 0 | 0 | -50006.63 | ***0.00*** |
| 0 | 1 | 0 | 0 | 0 | 0 | 0 | 0 | 0 | 0 | 0 | 0 | 0 | 0 | 0 | 0 | 0 | 1 | 0 | 0 | 0 | 0 | 0 | 0 | 0 | 0 | 0 | 0 | 0 | -50004.53 | ***2.10*** |
| 0 | 0 | 1 | 0 | 0 | 0 | 0 | 0 | 0 | 0 | 0 | 0 | 0 | 0 | 0 | 0 | 0 | 0 | 0 | 0 | 0 | 0 | 0 | 0 | 0 | 0 | 0 | 0 | 0 | -50005.53 | ***1.10*** |
| **Cyclic loop – transformation of DNAoxoG** | | | | | | | | | | | | | | | | | | | | | | | | | | | | | | |
| 0 | 0 | 0 | 1 | 0 | 1 | 0 | 0 | 0 | 0 | 0 | 0 | 0 | 0 | 0 | 0 | 0 | 0 | 0 | 0 | 0 | 0 | 0 | 0 | 0 | 0 | 0 | 0 | 0 | -76808.08 | ***0.00*** |
| 0 | 0 | 0 | 0 | 1 | 1 | 0 | 0 | 0 | 0 | 0 | 0 | 0 | 0 | 0 | 0 | 0 | 1 | 0 | 0 | 0 | 0 | 0 | 0 | 0 | 0 | 0 | 0 | 0 | -76807.18 | ***0.90*** |
| 0 | 0 | 0 | 1 | 0 | 0 | 1 | 0 | 0 | 0 | 0 | 0 | 0 | 0 | 0 | 0 | 0 | 1 | 0 | 0 | 0 | 0 | 0 | 0 | 0 | 0 | 0 | 0 | 0 | -76807.48 | ***0.60*** |
| 0 | 0 | 0 | 1 | 0 | 0 | 0 | 1 | 0 | 0 | 0 | 0 | 0 | 0 | 0 | 0 | 1 | 1 | 0 | 0 | 0 | 0 | 0 | 0 | 0 | 0 | 0 | 0 | 0 | -76807.51 | ***0.57*** |
| **Lp transformations** | | | | | | | | | | | | | | | | | | | | | | | | | | | | | | |
| 0 | 0 | 1 | 0 | 0 | 0 | 0 | 0 | 0 | 0 | 0 | 1 | 0 | 0 | 0 | 0 | 1 | 0 | 0 | 0 | 0 | 0 | 0 | 0 | 0 | 0 | 0 | 0 | 0 | -74735.11 | ***0.00*** |
| 1 | 0 | 0 | 0 | 0 | 0 | 0 | 0 | 0 | 0 | 0 | 0 | 1 | 0 | 0 | 0 | 1 | 0 | 0 | 0 | 0 | 0 | 0 | 0 | 0 | 0 | 0 | 0 | 0 | -74734.57 | ***0.55*** |
| 1 | 0 | 0 | 0 | 0 | 0 | 0 | 0 | 0 | 0 | 0 | 0 | 0 | 1 | 1 | 0 | 0 | 0 | 0 | 0 | 0 | 0 | 0 | 0 | 0 | 0 | 0 | 0 | 0 | -74735.17 | ***-0.05*** |
| 1 | 0 | 0 | 0 | 0 | 0 | 0 | 0 | -1 | 1 | 0 | 0 | 0 | 1 | 0 | 1 | 0 | 0 | 0 | 0 | 0 | 0 | 0 | 0 | 0 | 0 | 0 | 0 | 0 | -74735.87 | ***-0.76*** |
| **Hydrogen donor reconstritution** | | | | | | | | | | | | | | | | | | | | | | | | | | | | | | |
| 0 | 0 | 0 | 0 | 0 | 0 | 0 | 0 | 0 | 2 | 0 | 0 | 0 | 0 | 0 | 0 | 0 | 0 | 0 | 0 | 0 | 0 | 0 | 0 | 0 | 0 | 0 | 0 | 0 | -28073.57 | ***0.00*** |
| 0 | 0 | 0 | 0 | 0 | 0 | 0 | 0 | 1 | 0 | 1 | 0 | 0 | 0 | 0 | 0 | 0 | 0 | 0 | 0 | 0 | 0 | 0 | 0 | 0 | 0 | 0 | 0 | 0 | -28075.52 | ***-1.94*** |

***Fig. SI.2.2*** *The equation system from Fig.2.1 is here used to determine the free energies of selected processes. The upstream and downstream processes are discussed in Figs. 4.1.5 and 4.1.6, while the free energies for each reaction in the fatty acid production network in Fig. 4.1.1 are taken from the lower four reactions.*

## SI.3 Container self-assembly energetics

At equilibrium, the chemical potentials of decanoic acid (DA), $\mu_{\mathrm{uni}}^{(\mathrm{DA})}$, decanoate (D⁻), $\mu_{\mathrm{uni}}^{(\mathrm{D}^-)}$, and proton ($\mathrm{H}^+$), $\mu_{w}^{(\mathrm{H}^+)}$, satisfy the relationship

$$\mu_{\mathrm{uni}}^{(\mathrm{DA})} = \mu_{\mathrm{uni}}^{(\mathrm{D}^-)} + \mu_{w}^{(\mathrm{H}^+)}.$$

The chemical potentials of DA and D⁻ in the unimer state, $\mu_{\mathrm{uni}}^{(\mathrm{DA})}$and $\mu_{\mathrm{uni}}^{(D^-)}$, are expressed in terms of their standard chemical potentials, $\mu_{\mathrm{uni}}^{0\,(\mathrm{DA})}$and $\mu_{\mathrm{uni}}^{0\,(\mathrm{D}^-)}$, and their molar concentrations, $c_{\mathrm{uni}}^{(\mathrm{DA})}$ and $c_{\mathrm{uni}}^{(\mathrm{D}^-)}$, respectively, as

$$\mu_{\mathrm{uni}}^{(DA)} = \mu_{\mathrm{uni}}^{0\,(DA)} + k_B T \ln\,(c_{\mathrm{uni}}^{(DA)}/c^0)$$

$$\mu_{\mathrm{uni}}^{(D^-)} = \mu_{\mathrm{uni}}^{0\,(D^-)} + k_B T \ln\,(c_{\mathrm{uni}}^{(D^-)}/c^0) - e\psi_{\mathrm{w}},$$

where the standard concentration $c^0 = 1\mathrm{M}$. Here, $e\psi_{\mathrm{w}}$ is the electrostatic potential energy of the ionic molecule, where $\psi_{\mathrm{w}}$ is the electrostatic potential in water phase and $e$ is the elementary charge ($1.602\times 10^{-19}$ C). The chemical potential of proton is expressed in terms of its molar concentrations, $c_{w}^{(H^+)}$, as

$$\mu_{w}^{(H^+)} = k_B T \ln\,(c_{w}^{(H^+)}/c^0) + e\psi_{w}$$

where the standard chemical potential of protons is set as $\mu_{H^+}^{0} = 0$. Then, we obtain

$$\mu_{\mathrm{uni}}^{0\,(\mathrm{DA})} + k_B T \ln\,(c_{\mathrm{uni}}^{(DA)}/c^0) = \mu_{\mathrm{uni}}^{0\,(\mathrm{D}^-)} + k_B T \ln\,(c_{\mathrm{uni}}^{(D^-)}/c^0) - e\psi_{\mathrm{w}} + k_B T \ln(c_{w}^{(H^+)}/c^0) + e\psi_{\mathrm{w}}$$

The ratio of molar concentrations of decanoate to decanoic acid in the unimer state, $c_{\mathrm{uni}}^{(D^-)}/c_{\mathrm{uni}}^{(DA)}$, is determined by the acid dissociation constant $K_a = 1.3\times 10^{-5}$ ($pK_a = 4.9$) and the solution pH, as

$$c_{\mathrm{uni}}^{(D^-)}/c_{\mathrm{uni}}^{(DA)} = 10^{\mathrm{pH}-pK_a}$$

For example, at pH 8 (assumption), this ratio becomes $10^{3.1}$, indicating that decanoate exists predominantly in the solution. In the present system, DA and D⁻ in the unimer state coexist with DA/ D⁻ vesicles, and the concentration of D⁻ in solution is essentially equal to the CVC. Under these conditions, D⁻ in the unimer state is in equilibrium with D⁻ within the vesicle membrane.

We next evaluate the chemical potentials of DA and D⁻ within the vesicle membrane.
Let the number of DA molecules in a vesicle be $M$, with a standard chemical potential $\mu_{\mathrm{ves}}^{0\,(DA)}$, the number of D⁻ molecules be $N$, with a standard chemical potential $\mu_{\mathrm{ves}}^{0\,(D^-)}$, and the molar concentration of vesicles in solution be $c_{\mathrm{ves}}$. Then, the chemical potential of one vesicle is expressed as

$$\mu_{\mathrm{ves}} = M\mu_{\mathrm{ves}}^{0\,(DA)} + N\mu_{\mathrm{ves}}^{0\,(\mathrm{D}^-)} + k_B T[M\,\ln x_{\mathrm{ves}}^{DA} + N \ln x_{\mathrm{ves}}^{\mathrm{D}^-}] + k_B T \ln(c_{\mathrm{ves}}/c^0) - Ne\psi_{\mathrm{ves}}$$

where $x_{\mathrm{ves}}^{DA} = M/(M+N)$ is the mole fraction of DA in the membrane, $x_{\mathrm{ves}}^{\mathrm{D}^-} = N/(M+N)$ is that of D⁻, $\psi_{\mathrm{ves}}$ is the electrostatic potential at vesicle membrane. The first and second terms represent the standard chemical potentials of DA and D⁻ in the vesicle membrane, the third term accounts for the entropy of mixing, the fourth term corresponds to the translational entropy of vesicles, and the fifth term is the electrostatic potential contribution. However, the vesicle chemical potential can also be written as

$$\mu_{\text{ves}} = M\mu_{\text{ves}}^{(DA)} + N\mu_{\text{ves}}^{(\text{D}^-)}$$

Thus, the chemical potentials of DA and D⁻ in the vesicle membrane are

$$\mu_{\text{ves}}^{(DA)} = \mu_{\text{ves}}^{0\,(DA)} + k_B T \ln\frac{M}{M+N} + \frac{k_B T}{M+N}\ln(c_{\text{ves}}/c^0)$$

$$\mu_{\text{ves}}^{(D^-)} = \mu_{\text{ves}}^{0\,(D^-)} + k_B T \ln\frac{N}{M+N} + \frac{k_B T}{M+N}\ln(c_{\text{ves}}/c^0) - e\psi_{\text{ves}}$$

Here, the translational entropy term has been equally distributed across the molecules. If the vesicle membrane pH is $\text{pH}_{\text{ves}}$ and the equilibrium constant between DA and D⁻ in the membrane is $K_{\text{ves}}$, then the ratio of molecule numbers is given by

$$\frac{N}{M} = \frac{K_{\text{ves}}}{10^{-\text{pH}_{\text{ves}}}}$$

For fatty acid vesicles, the ratio is approximately 1:1, *i.e.*, $N = M$. In this case,

$$\mu_{\text{ves}}^{(DA)} = \mu_{\text{ves}}^{0\,(DA)} + k_B T \ln\frac{1}{2} + \frac{k_B T}{2N}\ln(c_{\text{ves}}/c^0)$$

$$\mu_{\text{ves}}^{(D^-)} = \mu_{\text{ves}}^{0\,(D^-)} + k_B T \ln\frac{1}{2} + \frac{k_B T}{2N}\ln(c_{\text{ves}}/c^0) - e\psi_{\text{ves}}$$

The molar concentrations of DA and D⁻ in the vesicle membrane are then $c_{\text{ves}}^{(DA)} = Nc_{\text{ves}}$, $c_{\text{ves}}^{(\text{D}^-)} = Nc_{\text{ves}}$. With $c^0 = 1\ M$, the chemical potentials can be rewritten as

$$\mu_{\text{ves}}^{(DA)} = \mu_{\text{ves}}^{0\,(DA)} + k_B T \ln\frac{1}{2} + \frac{k_B T}{2N}\ln(c_{\text{ves}}^{(DA)}/N)$$

$$\mu_{\text{ves}}^{(\text{D}^-)} = \mu_{\text{ves}}^{0\,(\text{D}^-)} + k_B T \ln\frac{1}{2} + \frac{k_B T}{2N}\ln(c_{\text{ves}}^{(\text{D}^-)}/N) - e\psi_{\text{ves}}$$

In the following, we focus on D⁻. At equilibrium,

$$\mu_{\text{uni}}^{(\text{D}^-)} = \mu_{\text{ves}}^{(\text{D}^-)}$$

or explicitly,

$$\mu_{\text{uni}}^{0\,(\text{D}^-)} + k_B T \ln\left(c_{\text{uni}}^{(\text{D}^-)}\right) - e\psi_{\text{uni}} = \mu_{\text{ves}}^{0\,(\text{D}^-)} + k_B T \ln\frac{1}{2} + \frac{k_B T}{2N}\ln(c_{\text{ves}}^{(\text{D}^-)}/N) - e\psi_{\text{ves}}$$

$$\mu_{\text{uni}}^{0\,(\text{D}^-)} - \mu_{\text{ves}}^{0\,(\text{D}^-)} = k_B T \ln\left[\frac{1}{2c_{\text{uni}}^{(\text{D}^-)}}\left(\frac{c_{\text{ves}}^{(\text{D}^-)}}{N}\right)^{\frac{1}{2N}}\right] - e\Delta\psi$$

where $\Delta\psi = \psi_{\text{ves}} - \psi_{\text{uni}}$ is the surface potential. Approximating the surface potential using the measured zeta potential of fatty acid vesicles in pure water (−62 mV), we have

$$e\Delta\psi \approx -k_B T \ln 10$$

Thus,

$$\mu_{\mathrm{uni}}^{0\,(\mathrm{D}^-)} - \mu_{\mathrm{ves}}^{0\,(\mathrm{D}^-)} = k_B T \ln\left[\frac{10}{2c_{\mathrm{uni}}^{(\mathrm{D}^-)}}\left(\frac{c_{\mathrm{ves}}^{(\mathrm{D}^-)}}{N}\right)^{\frac{1}{2N}}\right]$$

$$c_{\mathrm{ves}}^{(\mathrm{D}^-)} = N\left\{\frac{c_{\mathrm{uni}}^{(\mathrm{D}^-)}}{5}\exp\left(\frac{\mu_{\mathrm{uni}}^{0\,(\mathrm{D}^-)} - \mu_{\mathrm{ves}}^{0\,(\mathrm{D}^-)}}{k_B T}\right)\right\}^{2N}$$

The critical vesicle concentration (CVC) corresponds to the condition

$$\frac{c_{\mathrm{CVC}}^{(\mathrm{D}^-)}}{5} = \exp\left(-\frac{\mu_{\mathrm{uni}}^{0\,(\mathrm{D}^-)} - \mu_{\mathrm{ves}}^{0\,(\mathrm{D}^-)}}{k_B T}\right)$$

Experimentally, the CVC of $\mathrm{D}^-$ is found to be 57 mM. Then, we obtain

$$\mu_{\mathrm{uni}}^{0\,(\mathrm{D}^-)} - \mu_{\mathrm{ves}}^{0\,(\mathrm{D}^-)} = 4.5 k_B T = 0.12\mathrm{eV}.$$

Here, $\mu_{\mathrm{uni}}^{0\,(\mathrm{D}^-)}$ and $\mu_{\mathrm{ves}}^{0\,(\mathrm{D}^-)}$ denote the standard chemical potentials of $\mathrm{D}^-$ in the unimer and vesicle membrane states, respectively, which are equivalent to the standard Gibbs free energies of formation: $\mu_{\mathrm{uni}}^{0\,(\mathrm{D}^-)} = \Delta G_f^0[\mathrm{D}_{\mathrm{uni}}^-]$, $\mu_{\mathrm{ves}}^{0\,(\mathrm{D}^-)} = \Delta G_f^0[\mathrm{D}_{\mathrm{ves}}^-]$. This difference thus represents the standard Gibbs free energy difference between $\mathrm{D}^-$ in the unimer state and in the vesicle membrane state.